\documentclass[epj]{SVJour}
\usepackage{amssymb,epsf}

\newcommand{\bx}{{\bf x}}
\newcommand{\bq}{{\bf q}}

\newcommand{\bk}{{\bf k}}

\newcommand{\bQ}{{\bf Q}}
\newcommand{\bN}{{\bf 0}}

\newcommand{\tR}{\tilde R}
\newcommand{\tV}{\tilde V_L}

\newcommand{\cH}{{\cal H}}
\newcommand{\bbox}[1]{\mbox{\boldmath ${#1}$}}
\newcommand{\bphi}{\bbox{\phi}}

\begin{document}

\title{Disorder Driven Roughening Transitions of Elastic Manifolds
  and Periodic Elastic Media}

\dedication{Dedicated to Franz Schwabl on the occasion of 
  his 60th birthday.}

\author{Thorsten Emig 
  \thanks{Email address: {\tt te@thp.uni-koeln.de}}
  \and 
  Thomas Nattermann 
}

\institute{
  Institut f\"ur Theoretische Physik, Universit\"at zu K\"oln,
  Z\"ulpicher Str.\ 77, D-50937 K\"oln  
}

\date{\today}

\abstract{The simultaneous effect of both disorder and crystal-lattice
  pinning on the equilibrium behavior of oriented elastic objects is
  studied using scaling arguments and a functional renormalization
  group technique. Our analysis applies to elastic manifolds, e.g.,
  interfaces, as well as to periodic elastic media, e.g.,
  charge-density waves or flux-line lattices. The competition between
  both pinning mechanisms leads to a continuous, disorder driven
  roughening transition between a flat state where the mean relative
  displacement saturates on large scales and a rough state with
  diverging relative displacement. The transition can be approached by
  changing the impurity concentration or, indirectly, by tuning the
  temperature since the pinning strengths of the random and crystal
  potential have in general a different temperature dependence.  For
  $D$ dimensional elastic manifolds interacting with either
  random-field or random-bond disorder a transition exists for
  $2<D<4$, and the critical exponents are obtained to lowest order in
  $\epsilon=4-D$.  At the transition, the manifolds show a
  superuniversal logarithmic roughness.  Dipolar interactions render
  lattice effects relevant also in the physical case of $D=2$. For
  periodic elastic media, a roughening transition exists only if the
  ratio $p$ of the periodicities of the medium and the crystal lattice
  exceeds the critical value $p_c=6/\pi\sqrt{\epsilon}$. For $p<p_c$
  the medium is always flat.  Critical exponents are calculated in a
  double expansion in $\mu=p^2/p_c^2-1$ and $\epsilon=4-D$ and fulfill
  the scaling relations of random field models.}

\PACS{
{68.35.Ct}  {Interface structure and roughness} 
\and
{71.45.Lr} {Charge-density-wave systems}
\and
{64.70.Rh} {Commensurate-incommensurate transitions}
\and
{68.35.Rh} {Phase transitions and critical phenomena}
\and
{05.20.-y} {Statistical mechanics}
\and
{74.60.Ge}  {Flux pinning, flux creep, and flux-line lattice dynamics}
}

\authorrunning{T.~Emig, T.~Nattermann}
\titlerunning{Disorder Driven Roughening Transitions ...}
\maketitle

\section{Introduction}

The thermal roughening transition (RT) from a faceted to a smooth
surface of a crystal is one of the paradigms of condensed matter
physics \cite{Chaikin+95}.  It has been observed directly for a number
of materials, in particular on surfaces of crystalline Helium 4
\cite{Wolf+85}.  A similar transition occurs between a rough
delocalized and a flat localized state of the interface in the three
dimensional Ising and related lattice models
\cite{Weeks80,vanBeijeren+87}.  At the RT temperature $T_R$ the free
energy of a step on the surface vanishes.  Apart from its shape also
other physical properties are influenced by the presence of the RT,
e.g.,  the growth (drift) velocity ${u}$ of the crystal surface
(interface) changes dramatically at the RT from ${u} \sim \exp(-C/f)$
for $T<T_R$ to ${u}\sim f$ for $T>T_R$ \cite{Wolf+85}. Here $f$
denotes the driving force density, which is proportional to the
difference $\Delta \mu$ of the chemical potentials of the crystal and
its melt in the case of crystal growth and the magnetic field in the
case of Ising magnets, respectively.

A third closely related example for the RTs are the lock-in
transitions of periodically modulated structures into an underlying
crystalline matrix.  Examples for the modulated structures are spin or
charge density waves (CDWs) in anisotropic metals
\cite{Frohlich54,Gruner94}, mass density waves in superionic
conductors \cite{Bak+82}, polarization density waves in incommensurate
ferrolectrics \cite{Dvorak79}, Wigner crystals \cite{Andrei+88},
magnetic bubble arrays \cite{Seshadri+92}, flux line lattices (FLLs)
\cite{Balents+95b,Nieber+93} and nematic elastomers \cite{Fridrikh+97},
etc. The phenomenology of these bulk systems is even more rich than
that of sur- or interfaces, since in general a large number of
different locked-in structures separated by incommensurate phases may
occur.

Common feature of the examples mentioned is the existence of a
$D$-dimensional deform-able object (i.e. $D=1$ for single flux lines,
$D=2$ for surfaces or interfaces, $D=3$ for CDWs or the FLL etc.) the
energy of which - in the absence of lattice effects - can be described
by elasticity theory.  The distortion of the elastic object is
expressed in general by an $N$-component displacement field
$\phi^i({\bf x}), i=1...N$, where $N=1$ for surfaces, interfaces and
CDWs, $N=2$ for flux lines or FLLs etc. These elastic objects
interact with the potential of the underlying crystalline matrix,
which favors (in a commensurate situation) certain spatially
homogeneous ("flat") configurations because of their lower energy $E$.
On the other hand, thermal fluctuations favor a spatially
inhomogeneous ("rough") configuration because these have the higher
entropy $S$. The real macroscopic configuration is that of the lowest
free energy $F=E-TS$, from which we expect the occurrence of a
roughening transition at a temperature $T=T_R$.

Later we will subdivide these elastic objects into {\it manifolds},
which comprise single flux lines or domain walls, and {\it periodic
  media}, i.e. periodic lattices of flux lines or domain walls, CDWs
etc. This distinction will become necessary because manifolds and
periodic media couple in different ways to the disorder.  However, a
distinction is unnecessary as long as we consider merely thermal
fluctuations.

The interest in the investigation of the roughening transition goes
also beyond its immediate applications in condensed matter physics.
In computer simulations of condensed matter systems a {\it fake}
periodic lattice potential - in addition to that of the crystalline
matrix - appears frequently in the form of a grid on which the
simulations are done, which has nothing to do with the physics of the
problem. In order to retrieve physically relevant informations from
these simulations one has to make sure to be above the roughening
transition of this fake potential. This remark applies also to lattice
gauge theories in high energy physics \cite{Wallace80,Montvay+94}.

Besides of its frequent occurrence the second interesting aspect of
the thermal RT is its {\it universality}.  Although the RT temperature
$T_R$ itself is non-universal and depends in particular on the elastic
properties of the object under consideration (e.g. on the orientation
of a vicinal surface relative to the main crystalline axises), the
universal features like the singular behavior of the thermodynamic
quantities are the same for all models with the same spatial dimension
$D$ and number of field components $N$.  Moreover, the thermal RT for
$D=2$-dimensional surfaces was shown to be in the same universality
class as the metal-insulator transition of a 2D-Coulomb gas
\cite{Chui+76}, the Kosterlitz-Thouless transition of a 2D XY-model
\cite{Knops77} (both with an inverted temperature axis), the phase
transition of the 2D sine-Gordon model \cite{Jose76} and the
transition of the 1D XXZ-chain at T=0 \cite{Luther+75}.

$D=2$ is indeed the upper critical dimension for the thermal RT since
$D> 2$ dimensional manifolds are flat even without a lattice
potential.

Interesting but less known is the fact that Kosterlitz
\cite{Kosterlitz77} and Forgacs {\it et al.} \cite{Forgacs+91}
considered the thermal RT for $1< D\le 2$ interface dimensions and found
a superuniversal (i.e. dimension independent) logarithmic roughness
{\it at} the transition.  Finally, the thermal RT disappears in $D=1$
dimensions: $1$-dimensional interfaces are rough at all finite T
\cite{FisherDS+82}.

{\it Disorder} is an even more efficient source of roughening of
manifolds than thermal fluctuations and physically important as well
\cite{Halpin+95}.

Roughening of interfaces due to disorder was considered to determine
the lower critical dimension of the random field Ising model
\cite{Villain82,Grinstein+82,Binder83} and the mobility of domain
walls in disordered magnets
\cite{Villain84,Ioffe+87,Nattermann87,Nattermann+90,Nattermann+88b,Nattermann+92}.
It was argued that in the presence of disorder interfaces of
dimensions $D\leq 2$ are always rough, but undergo a roughening
transition as a function of the disorder strength for $D > 2$
dimensions \cite{Nattermann84,Nattermann85}.  Bouchaud and Georges
\cite{Bouchaud+92b} considered explicitly the competition between
lattice and impurity pinning in $2\leq D\leq 4$ dimensions using a
variational calculation.  Surprisingly they found {\it three} phases:
a weakly disordered flat, a rough glassy and an intermediate flat
phase with strong glassy behavior.  For $D\le 2$ only the glassy rough
phase was expected to survive, but that cross-over regions of the two
other phases would still be seen in the short scale behavior. However,
it should be kept in mind that the variational calculation is an
uncontrolled approximation and typically reproduces the results
obtained from Flory-like arguments.

It is the aim of the present paper to give a unified renormalization
group description of the roughening transition of elastic manifolds
and periodic media in the presence of quenched disorder.  Some
preliminary results were already published \cite{Emig+97,Emig+98}.  In
this paper we focus on the derivation of the renormalization group
flow equations and their solutions for the different cases. To make
the paper more self-contained some of the results reported already in
Refs. \cite{Emig+97,Emig+98} are included here as well.  The physics
is complicated by the occurrence of various input length scales like
the periodicity of the crystalline matrix $a_0/p$, the correlation
length of the disorder $\xi_0$ and the periodicity of the periodic
medium $l$.

The outline of the paper is as follows. In Sec.~2 the model for the
pinned elastic objects is described. The different classes of disorder
correlations are briefly reviewed. Section 3 analyses the model on the
basis of simple scaling arguments. A qualitative picture of the phase
diagram is developed and the characteristic length scales are
identified. In Section 4 we proceed with the derivation of the
functional RG equations to lowest order in $D=4-\epsilon$ dimensions.
The functional RG flow of the disorder correlator is reduced to a flow
of single parameters in Section 5. The fixed points of this RG flow
are analyzed for vanishing lattice potential, reproducing the known
results. The effect of a finite lattice potential is considered in
Section 6. The critical exponents of the resulting roughening
transition are calculated to lowest order in $\epsilon$ for the
different types of disorder. Experimental implications of our results
and related recent experiments are discussed in Section 7.  The last
section concludes with a summary and discussion of our results and
briefly comments on related open questions not treated in this paper.

\section{The Model}

As mentioned already in the Introduction, the elastic objects are
described by an $N$-component phase field ${\bphi}({\bf x})$ with
components $\phi^i({\bf x})$, $i=1,\ldots, N$, which interacts both
with an underlying periodic crystalline lattice potential $V_L(\bphi)$
and a random potential $V_R(\bphi,\bx)$ specified below.  The total
Hamiltonian is then given by
\begin{equation}
  \label{genHam}
  \cH=\gamma \int d^D \bx \left\{ \frac{1}{2} \bbox{\nabla \phi} \cdot
    \bbox{\nabla \phi} - V_L(\bphi(\bx)) - V_R(\bphi(\bx),\bx) \right\}.
\end{equation}
The strength of the pinning potentials is measured in units of the
elastic stiffness constant $\gamma$ which controls isotropic elastic
deformations. More complicated elastic moduli could be considered but
turn out to be unessential as long as we are interested in universal
properties.  For manifold models, the elastic object is embedded into
a larger space of dimension $d=D+N$ whereas for periodic media the
dimension of the space in which the elastic structure is embedded is
equal to the internal dimension, $d=D$. The different realizations of
the model (\ref{genHam}) are summarized in Table \ref{tab_models}.

The periodic potential  $V_L(\bphi)$ can be rewritten as 
\begin{equation}
  \label{defL}
  V_L(\bphi)=\sum_{i=1}^N V(\phi^i).
\end{equation}
if the original crystalline matrix is assumed to have the symmetry of
an $N$--dimensional hyper--cubic lattice with spacing $a_0$.  Below we
will measure all length scales in units of $a_0$, i.e. we will put
$a_0\equiv 1$. The resulting $V(\phi^i)$ is then a periodic function
of periodicity $2\pi/p$.  The random pinning potential
$V_R(\bphi,\bx)$ is assumed to be Gaussian distributed with a
vanishing average $\overline{V_R(\bphi,\bx)}=0$ and two-point
correlation function
\begin{equation}
  \label{def_R}
  \overline{V_R(\bphi,\bx)V_R(\bphi',\bx')}=R(\bphi-\bphi')
  \delta^{(D)}(\bx-\bx').
\end{equation}

\begin{table*}
\caption{Different physical realizations of the model (\ref{genHam});
  topological defects are excluded.}
\label{tab_models}   
\begin{center}
\begin{tabular}{l|cccccc}
\hline\noalign{\smallskip}
Elastic object & $d$ & $D$ & $N$ & $\bphi$ & $p$ & $R_0(\bphi)$, 
$|\bphi|\gg \xi_0$\\
\noalign{\smallskip}\hline\noalign{\smallskip}
Interface \cite{Villain82,Grinstein+82,Binder83}
& $\ge 2$ & $d-1$ & $ 1 $ & $2\pi u/a$ & 1 & 
$-|\bphi |$ (random field)  \\
& & & & $u$ displacement & & \\
& & & & & & $e^{-\bphi^2/\xi_0^2}$ (random bond) \\
& & & & & & \\
density wave \cite{Gruner94,Bak+82,Dvorak79} 
& $\ge 2$ & $d$ & 1 & $\delta\rho ({\bf x})=\rho_1\cos{({\bf q}
{\bf x}+\bphi )}$ & $\bq=\bQ/p$ & $\cos{\phi}$  \\
& & & & & & \\
$XY$--magnet & & & & & & \\
in crystal and  & $\ge 2$ & $d$ & 1 & $\arctan(S_y/S_x)$ & $\ge 1$, number 
& $\cos{\phi}$  \\
random field & & & & & of easy directions& \\
& & & & & & \\
flux line lattice \cite{Balents+95b,Nieber+93} 
& $\ge 2$ & $d$ & 2 & $2\pi{\bf u}/l$ &
$p=l/a$ & $\sum_{\bQ\neq {\bf 0}} R_\bQ \cos(\bQ\bphi)$\\
& & & & $l$ flux line spacing & & $\bQ$ reciprocal lattice vectors\\
& & & & & & \\
$^4$He surface & $\ge 2$ & $d-1$ & 1 & $2\pi h/a$ & 1 & $-|\phi|$\\
in aerogel & & & & $h$ height variable & & \\
& & & & & & \\
nematic elastomer & $\ge 2$ & $d$ & 1 & $\arctan(n_y/n_x)$ & 1 & $\cos\phi$\\
in external and & & & & ${\bf n}$ unit director field & & \\
random field \cite{Fridrikh+97} & & & & & &\\
\noalign{\smallskip}\hline
\end{tabular}
\end{center}
\end{table*}
The starting point for our further calculations is the replica
Hamiltonian $\cH_n$, which follows, e.g., from the consideration of
the disorder averaged free energy
\begin{equation}
  \overline{F}=-T\lim_{n\to 0}\frac{1}{n}\left( {\rm Tr}e^{-\cH_n}-1\right)
  \label{free-energy}
\end{equation}
with
\begin{eqnarray}
  \cH_n=\frac{\gamma}{T}\sum\limits_{\alpha\beta =1}^n\int d^Dx & &\left\{
  \left[\frac{1}{2}\bbox{\nabla \phi}_{\alpha}\cdot\bbox{\nabla \phi}_{\beta}
  -V_L(\bphi_{\alpha})\right]\delta_{\alpha\beta}\right.\nonumber\\
  & & -\frac{\gamma}{2T}R(\bphi_{\alpha}-\bphi_{\beta})\bigg\}
  \label{FH_n}
\end{eqnarray}
The actual form of the disorder correlator $R(\bphi)$ depends on the
type of disorder and on the model under consideration.

We give here a brief discussion of the main classes of disorder
correlators:

For {\it elastic manifolds} one has to distinguish between random
field and random bond systems.

(i) For {\it random field system} \cite{Villain82,Grinstein+82} the bare
disorder correlator is given by
\begin{equation}
  R_0(\bphi)=-\Delta_0\left\{
  \begin{array}{lcl}
  |\bphi|\xi_0 & \mbox{for} & |\bphi| \gg \xi_0\\
  \bphi^2 & \mbox{for} & |\bphi| \ll \xi_0\\
  \end{array}\right. .
  \label{(i)}
\end{equation}
Here $\sqrt{\Delta_0}$ denotes the strength of the random field,
$\Delta_0$ is proportional to the concentration $n_{\rm imp}$ 
of the random field impurities.
$\xi_0$ is of the order of the maximum of the intrinsic domain wall
width and the correlation length of the random field.
The result (\ref{(i)}) was derived in \cite{Villain82,Grinstein+82} for
$N=1$ component fields. Its formal generalization to arbitrary $N>1$ is
straightforward but without direct physical application. 
However, this case will be included here
in order to enable the comparison of our results with that of the
variational approach of Bouchaud and Georges \cite{Bouchaud+92b}
which is expected to become exact for $N\to \infty$.

As an other interesting realization of a sine-Gordon model with random
field disorder, we suggest surfaces of crystalline $^4$He in an
aerogel. Here the silica strands act as quenched impurities on the
crystal surface. Fractal correlations in the aerogel density appearing
on short scales should be negligible if the correlation length
$\xi_\|$ of the surface exceeds all structural length scales of the
aerogel.

(ii) For {\it random bond systems} with
short range correlations in the disorder 
the bare disorder correlator is given by
\begin{equation}
  \label{R_rb}
  R_0(\bphi)=\Delta_0\xi_0^2\exp(-\bphi^2/2\xi_0^2).
\end{equation}
The length $\xi_0$ has the same meaning here as for random field
systems. Note that in our notation $\xi_0^2$ includes a factor $N$
such that $\xi_0^2/N$ remains finite in the limit $N\to \infty$. 

(iii) In the case of {\it periodic elastic media} the form of the bare
disorder correlator is dominated by the periodicity of the elastic
medium.  To show this more explicitly we start, for simplicity, with a
scalar field $\phi$.  The periodically modulated density can be
expanded in a Fourier series
\begin{equation}
  \label{charge_density}
  \rho(\bx ,\phi )=\sum_{m=0}^\infty \rho_m \cos[m(\bq\bx +\phi(\bx))]
\end{equation}
where $\bq$ denotes the modulation vector of the locked--in phase.
The coupling to the disorder potential $U(\bx )$ is of the form
\begin{equation}
  V_R(\phi ,\bx )=U(\bx )\rho (\bx ,\phi )
  \label{V_R}
\end{equation}
where $U(\bx )=U_0\left(\sum_i\delta (\bx -\bx_i)-n_{\rm imp}\right)$.
Here $\bx_i$ and $n_{\rm imp}$ are the positions and the concentration
of the impurities. The disorder average is performed by integration
over all possible impurity positions,
\begin{equation}
  \overline{A} \equiv \prod\limits_{i}\int\frac{d^Dx_i}{\cal V}
  A(\{ x_i\}).
  \label{Aaverage}
\end{equation}
Here the integration extends over the volume ${\cal V}$ of the system.
With (\ref{def_R}), (\ref{charge_density}), (\ref{V_R}) and neglecting
rapidly oscillating terms we get then in a straightforward manner
\begin{equation}
  R_0(\phi)=\frac{1}{2} n_{\rm imp} U_0^2 \sum_{m=0}^\infty \rho_m^2 
  \cos(m\phi).
\end{equation}
This expression can be extended to vector fields $\bphi(\bx)$ as
discussed by Giamarchi and Le Doussal \cite{Giamarchi+94} in the
context of flux line arrays in superconductors, and gives
\begin{equation}
  R_0(\bphi)=\sum_{\bQ\neq {\bf 0}} R_\bQ \cos(\bQ\bphi)
\end{equation}
with Fourier coefficients $R_\bQ$ \cite{Giamarchi+95} where $\bQ$ are
the reciprocal lattice vectors of the periodic medium. Generally,
rewriting the periodic density $\rho(\bx,\bphi)$ in terms of a
displacement field $\bphi$, also gradient terms of the form $-\rho_0
\sum_{i=1}^N \partial_i\phi^i$ are generated in Eq.
(\ref{charge_density}) which lead to terms of the form
$\partial_i\phi^i_\alpha \partial_j\phi^j_\beta$ in Eq.  (\ref{FH_n}).
But these gradient terms can be shown to be strongly irrelevant for
the large distance behavior for $D>2$ \cite{Giamarchi+95}. Therefore,
in the following this interaction will be neglected.

\section{Scaling arguments}

In this section we present some elementary scaling considerations for the
system described by the Hamiltonian (\ref{genHam}), which give a qualitative 
picture of the phase diagram, the critical dimensions and the characteristic
length scales. Since we are interested only in qualitative
estimates, we neglect all numerical factors of order unity.

We start with the {\it elastic manifolds} and restrict our analysis
to the interface case $N=1$. The extension to $N>1$ does not
change the qualitative conclusions derived for $N=1$.
Since disorder fluctuations turn out to be much stronger than
those of thermal fluctuations, we put $T=0$ from the very beginning,
postponing the discussion of the influence of finite temperatures to the
end of this section.

\begin{figure}[htb]
\begin{center}
\leavevmode
\epsfxsize=0.7\linewidth
\epsfbox{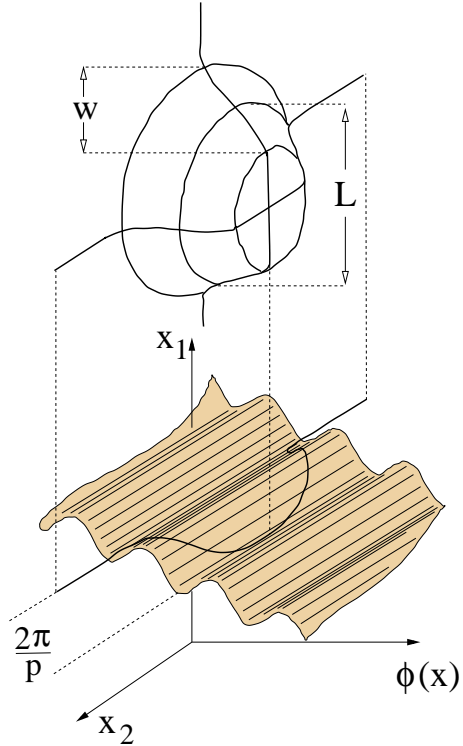}
\end{center}
\caption{Two dimensional manifold forming a terrace to gain energy
  from the random potential without any bulk energy cost from the
  periodic potential.}
\label{fig_terrace}
\end{figure}

In the absence of disorder, $\Delta_0=0$, the interface is flat.
Typical excitations consist of terraces of height $2\pi /p$ and linear
extension $L$. The terraces are bound by steps of width $w\approx
1/p\sqrt{v}$ and energy $E_{\rm step}\simeq \gamma\sqrt{v_0}L^{D-1}/p$
where we used for simplicity $V_L(\phi )\simeq v_0\cos{p\phi}$ for the
lattice potential, see Fig. \ref{fig_terrace}.  Disorder can lower the
energy of terraces since the locally displaced interface sees a
different random potential. To be specific, we find for the energy of
a hump of total height $\phi$ consisting of $\phi p/2\pi$ steps
\begin{equation}
  E_{\rm total}/\gamma\simeq \sqrt{v_0}L^{D-1}\phi-\sqrt{\Delta_0\xi_0
  L^{D}\phi}.
  \label{humpenergy}
\end{equation}
The second part represents the energy gain from the random
field. Minimizing the total energy (\ref{humpenergy}) with respect to
$\phi$ we obtain $p\phi\simeq (L/L_R)^{2-D}$ with
\begin{equation}
  L_R\simeq (v_0/\xi_0 \Delta_0p)^{1/(2-D)}
  \label{minL_R} 
\end{equation}
for $D<2$, and $L$ has to be replaced by the lattice spacing $a_0$ for
$D>2$.  In deriving (\ref{minL_R}) we implicitly assumed $\xi_0\ll
a_0/p$ with $a_0=1$ for the estimate of the disorder energy in
(\ref{humpenergy}). In the opposite limit $\xi_0\gg a_0/p$
perturbation energy gives for the disorder energy
$\sqrt{\Delta_0L^D}$.  A useful interpolation formula for $L_R$ is
therefore
\begin{equation}
  L_R\simeq\left(\frac{v_0(\xi_0p+1)}{\Delta_0\xi_0p}\right)^{1/(2-D)}
  \label{minL_R1}
\end{equation}
For $D<2$ the flat phase is unstable and the interface is therefore
rough on scales $L>L_R$ even in the presence of the crystal potential.
Since the energy cost for terraces vanishes now on length scales
$L\simeq L_R$, the system is described on larger scales by an
effective {\it elastic} Hamiltonian with a stiffness constant
$\gamma_{\rm eff}$. A
rough estimate for $\gamma_{\rm eff}$ follows from balancing the
elastic energy and the bare energy cost for a terrace of height
$2\pi /p$ on the length scale $L_R$,
\begin{equation}
  \gamma_{\rm eff}L_R^{D-2}\approx \gamma\sqrt{v_0}
  L_R^{D-1}p,
\end{equation}
which gives $\gamma_{\rm eff}\approx \gamma L_R/w$
\cite{Nattermann84,Nattermann85}. On scales $L\gg L_R$ the hump height
scales as $p\phi\simeq 2\pi (L/L_R)^\zeta$ where $\zeta=(4-D)/3$ is
the roughness exponent for random fields
\cite{Villain82,Grinstein+82}.  $L_R$ is physically meaningful only if
it is larger than the bare kink width, $L_R > w$. This condition can
be rewritten as
\begin{equation}
  v_0>v_{0,c}\approx \left(\frac{\Delta_0\xi_0p^{D-1}}{1+\xi_0p}
  \right)^{2/(4-D)}.
  \label{v0}
\end{equation}
If $v_0\to v_{0,c}+$, $\gamma_{\rm eff}$ changes into $\gamma$.
For $v_0<v_{0,c}$, the lattice potential can be neglected and
$v_0$ has to be replaced by $v_{0,c}$ in the above formulas. 

So far we have considered the case $D<2$. For $D=2$ the length scale
$L_R$ can be obtained from an argument which has been originally
applied by Morgenstern {\it et al.} \cite{Morgenstern+81} and Binder
\cite{Binder83} to the surface of an Imry-Ma domain in the two
dimensional random field Ising model. Consider again a terrace of
height $2\pi/p$ and size $L$ in the otherwise planar interface. As can
be seen from Eq. (\ref{humpenergy}), both the step energy and the
random energy scale proportional to $L$. To calculate $L_R$ one has to
consider the distribution of random energies $E_{\rm dis}$ of such a
terraces resulting from different disorder configurations. The mean
value of this random energies vanishes, but its variance increases
with $L$.  For large enough terraces, the distribution for $E_{\rm
  dis}$ becomes Gaussian,
\begin{equation}
  \label{s_distri}
  {\cal P}(E_{\rm dis})=\frac{1}
  {4\pi L\sqrt{\Delta_0\xi_0/p}}\exp\left( -\frac{1}{2}
\frac{E_{\rm dis}^2}{8\pi L^2\Delta_0\xi_0/p}\right).
\end{equation}
Since the total energy cost for a terrace of circular shape is given
by the sum of the elastic energy $E_{\rm el}=2\pi^2\sqrt{v_0}L/p$ and
$E_{\rm dis}$, the probability ${\cal P}_<$ that the total energy
becomes negative can be easily estimated to be
\begin{equation}
 {\cal P}_< = \int^{-E_{\rm el}}_{-\infty} dE_{\rm dis}\, 
{\cal P}(E_{\rm dis}).
\end{equation}
This integral evaluated in the limit $\Delta \to 0$ gives
\begin{equation}
  {\cal P}_< \approx
  \frac{1}{\pi^2}\sqrt{\frac{p\xi_0\Delta_0}
    {v_0}}\exp\left(-\frac{\pi^3}{4}
    \frac{v_0}{p\xi_0\Delta_0}\right).
\end{equation}
Hence the density ${\cal P}_< \approx (L_R)^{-2}$ of such terraces
tends exponentially to zero as $\Delta_0\to 0$. Neglecting algebraic
factors, terraces of arbitrary size appear spontaneously beyond the
exponentially large length scale
\begin{equation}
  \label{2D_L_R}
  L_R\simeq \exp\left(\frac{\pi^3}{8}
    \frac{v_0}{p\xi_0\Delta_0}\right)
\end{equation}
such that the interface becomes asymptotically rough.  Thus, the flat
phase is unstable to a weak random field in $D\le 2$ dimensions. 

For $D>2$, we have to expect from the previous considerations that the
interface becomes rough only for $v_0<v_{0,c}$, i.e., that there is a
RT as a function of $v_0$ (or $\Delta_0$) at $v_0\simeq v_{0,c}$.
Since one could object that the argument for the occurrence of a
critical value $v_{0,c}$ for $v_0$ is restricted to $D\le 2$, we give
here an additional argument for the existence of a roughening
transition for $D>2$ by comparing the pinning forces which originate
from the periodic and the impurity potential, respectively.  If there
is no disorder, an external force (density) has to overcome a
threshold value
\begin{equation}
  f_v\approx p\gamma v_0
  \label{fv}
\end{equation}
to depin the interface. On the other hand, for $v_0=0$ the threshold force 
from the impurity potential on a length scale $L$ is given by
\cite{Nattermann90}
\begin{equation}
  f_{\Delta}(L)\approx\gamma\xi_0L_{\Delta}^{\zeta}L^{\zeta -2}
  \label{fDelta}
\end{equation}
where we introduced the Larkin--Fukuyama--Lee length
$L_{\Delta}\approx (\xi_0^2/\Delta_0)^{1/(4-D)}$
and the roughening exponent $\zeta$ of the manifold in the absence
of lattice pinning. The transition between the lattice pinned flat
and the impurity pinned rough phase occurs if both force
densities are equal on the length scale of the step width $w$. This gives
\begin{equation}
  v_{0,c}\approx p^{-2}(p\xi_0)^{2/\zeta}(\Delta_0/\xi_0^2)^{2/(4-D)}.
  \label{v0c}
\end{equation}
For random field systems, $\zeta =(4-D)/3$ if $p\xi_0\ll 1$
(asymptotic regime) and $\zeta =(4-D)/2$ if $p\xi_0\gg 1$
(perturbative regime), in complete agreement with (\ref{v0}).

For completeness we additionally show the stability of the rough phase
to a weak crystal potential. Indeed, lowest order perturbation theory
in $v_0$ yields an effective Hamiltonian where the periodic crystal
potential is replaced by a mass term $m^2 \phi^2/2$. Assuming a
Gaussian distributed field $\phi(\bx)$, which is justified to lowest
order in $\epsilon=4-D$, we obtain on length scales $L>L_\Delta$ for
the mass
\begin{eqnarray}
\label{pert}
  m^2&=&p^2v_0 \overline{\cos (p\phi)} = p^2 v_0 \exp[-p^2\overline{\phi^2}
  /2]\nonumber\\ 
  &=&p^2 v_0 \exp[-2\pi^2 (L/L_\Delta)^{2\zeta}].
\end{eqnarray}
Comparing this exponential behavior in $L$ with the $L^{2\zeta-2}$
scaling of the elastic energy, we conclude that the perturbation is
always irrelevant since $\zeta>0$ for the elastic manifold models.
Likewise, the flat phase is stable to a weak random potential as can
be shown easily by perturbation theory with respect to $V_R$.  From
the stability of both phases in $D>2$ dimensions we can expect that
there is a roughening transition for $D>2$ as a function of $v_0$ (or
$\Delta_0$) at $v_0\simeq v_{0,c}(\Delta_0)$.

In the rough phase, the mean square displacement of the interface can be 
estimated again from the energy
\begin{equation}
  \frac{E_{\rm total}}{\gamma L^{D-2}}\simeq\phi^2+\sqrt{v_0}L\phi
  e^{-p^2\phi^2/2}-\sqrt{\xi_0\Delta_0\phi}L^{(4-D)/2}.
  \label{EtotgammaL}
\end{equation}
Here we have included both the elastic and the step energy term to describe
the crossover from $v_0\ll v_{0,c}$ to $v_0\simeq v_c$. Moreover,
we have treated the lattice pinning term in a self consistent harmonic
approximation to account for a situation in which $p\phi\gtrsim 2\pi$.
In the rough phase, the hump height asymptotically scales with the
roughness exponent $\zeta=(4-D)/3$ as a
result of the balance between the first and the last term in
Eq. (\ref{EtotgammaL}). However, there is an intermediate length
scale region up to $\xi_\|$ in which the roughness of the interface is
dominated by the competition between the second and the last term of
Eq. (\ref{EtotgammaL}). In this region the roughness increases only
logarithmically,
\begin{equation}
  \overline{\phi^2}\approx \frac{2}{p^2}(D-2)
  \ln\left(\frac{L}{L_R}\right).
\end{equation}
This result will be confirmed below by the RG calculation which yields
also the exponent $\nu_\|$ of the crossover length $\xi_\|$ to first
order in $\epsilon=4-D$.  Finally, for $v_0\gg v_{0,c}$ the interface
is always flat. 

So far we have treated only the case of random fields. For {\it random
  bond systems} similar arguments apply but the asymptotic roughness
cannot be obtained correctly from a Flory argument.

For {\it periodic elastic media}, arguments identical to those used
before for manifolds give a lower critical dimension $D_c(=d_c)=2$ for
the occurrence of a roughening transition and the estimate (\ref{v0})
for the critical value of $v_{0,c}$ if we set $\xi_0p\gg 1$.

To estimate the influence of a weak crystal potential in the rough
phase, we compare again the scaling of the elastic energy with that of
the effective mass $m^2$ as given by Eq. (\ref{pert}). In what
follows, the logarithmic roughness of the structure,
$\overline{\phi^2}=\frac{\pi^2}{9}(4-D)\ln(L/L_\Delta)$, on length
scales $L>L_\Delta$ and for $2<D<4$, cf. Sec. \ref{sec5}, is crucial.
Due to this weak roughness, the effective mass decays only
algebraically on large length scales $L$,
\begin{equation}
  m^2=p^2 v_L \left(\frac{L}{L_\Delta}\right)^{-\pi^2 p^2 (4-D)/18}.
\end{equation}
The elastic energy simply scales as $L^{-2}$ and thus we conclude that
the crystal potential is a relevant perturbation if
\begin{equation}
  p<p_c\equiv \frac{6}{\pi\sqrt{4-D}}
\end{equation}
leading always to a flat phase.  For $p>p_c$ the rough phase is again
stable and one has to expect a roughening transition. Possible
subtleties about exchanging the limits of large $L$ and small
$\epsilon$ have been neglected here, but this result will be
reproduced below by a detailed RG double expansion in $\epsilon=4-D$
and $\mu=p^2/p_c^2-1$.

So far we have neglected thermal fluctuations. In $D>2$, where we have
anticipated the occurrence of a roughening transition, thermal
fluctuations are irrelevant. In $D\le 2$ fluctuations are still dominated 
by disorder fluctuations on large scales. However, for weak disorder
and sufficiently high temperatures $T>T_R$, a length cross--over
from thermal to disorder dominated pinning phenomena will occur as briefly
discussed in \cite{Nattermann84,Nattermann85}.

\section{Functional Renormalization Group}

Next we employ a functional RG calculation, which starts with the
disorder averaged replica-Hamiltonian (\ref{FH_n}).
We apply a momentum shell RG to this Hamiltonian, in which the
high-momentum modes of $\bphi(\bx)$ in a shell of infinitesimal width,
$|\bq| \in [\Lambda e^{-dl},\Lambda]$, are integrated out. The
momentum cutoff $\Lambda$ will be kept fixed by rescaling coordinates
and fields according to
\begin{equation}
  \bx=e^{dl} \bx', \quad \bphi(\bx)=e^{\zeta dl} \bphi'(\bx').
\end{equation}
Since the behavior of the elastic objects should be described by
zero-temperature fixed points, we allow the temperature to be
renormalized by keeping the elastic stiffness constant $\gamma$ fixed.
From the above scale changes we obtain the RG flow of the temperature,
\begin{equation}
\label{scal_T}
\frac{dT}{dl}=(2-D-2\zeta)T.
\end{equation}
It can be shown that this flow equation is exact for vanishing lattice
potential due to the tilt symmetry of the Hamiltonian (\ref{genHam})
for $V_L(\bphi)=0$ \cite{Schulz+88}. As we will see below, a finite
lattice potential leads to a renormalization of the elastic stiffness
which can be rewritten as a renormalization of the temperature. This
leads to an additional term in Eq. (\ref{scal_T}) but does {\it not}
change the temperature into a relevant RG variable.

The RG equations for $V_L(\bphi)$ and $R(\bphi)$ will be
derived in their functional form following the method developed by
Balents and Fisher \cite{Balents+93}. The two last terms of the
replica-Hamiltonian (\ref{FH_n}) will be considered as a small
perturbation ${\cal H}_{\rm pert}$ of the elastic gradient term
which we denote by $\cH_0$. To eliminate the high-momentum modes, the
field is divided into slowly and rapidly varying contributions,
$\bphi(\bx)=\bphi^<(\bx)+\bphi^>(\bx)$.  The feedback from integrating
out $\bphi^>(\bx)$ can be calculated by expanding the effective
replica-Hamiltonian for the slow fields $\bphi^<(\bx)$,
\begin{equation}
\tilde\cH_n=-\ln \int [d\bphi_\alpha^>]
 \exp(-\cH_n),
\end{equation}
in powers of ${\cal H}_{\rm pert}$. The first terms of this cumulant
expansion are given by
\begin{equation}
\label{cumulant}
  \tilde\cH_n=\cH_0+\langle \cH_{\rm pert} \rangle_{>,c}-\frac{1}{2}
\langle \cH_{\rm pert}^2 \rangle_{>,c}+\frac{1}{6}\langle 
\cH_{\rm pert}^3 \rangle_{>,c}-\ldots .
\end{equation}
Here $\langle\ldots\rangle_{>,c}$ denotes the connected expectation
value over $\bphi^>(\bx)$ with respect to $\cH_0$. Generally, new
interactions which are not present in the original Hamiltonian are
generated by the RG transformation and have to be included in ${\cal
  H}_{\rm pert}$ if they are relevant in the RG sense. We postpone the
discussion of such sort of terms to the step in the RG calculation
where they appear for the first time. 

Now we start to evaluate the series (\ref{cumulant}) order by order.
To obtain the feedback of {\it first order} from the perturbation, we
rewrite $\cH_{\rm pert}$ in its Fourier representation,
\begin{eqnarray}
\label{fourier_pert}
  -\cH_{\rm pert}&=&\frac{1}{2T^2}\sum_{\alpha\beta} \int_{\bx,
    \bk} e^{i\bk \bphi_{\alpha\beta}(\bx)}
  \tR(\bk)\nonumber\\
&+& \frac{1}{T}\sum_\alpha \int_{\bx,\bk} e^{i\bk
    \bphi_\alpha(\bx)}\tV(\bk), 
\end{eqnarray}
where we use here and in the following the vector $\bk$ for Fourier
transforms with respect to $\bphi$, whereas the vector $\bq$ will be
used to denote momenta in $\bx$ direction. Furthermore, we have
introduced the abbreviations $\int_{\bx,\bk}\equiv \int d^D \bx \int
d^N \bk /(2\pi)^N$, $\bphi_{\alpha\beta}(\bx)\equiv
\bphi_\alpha(\bx)-\bphi_\beta(\bx)$ and $\gamma/T \rightarrow 1/T$.

\begin{figure}[htb]
\begin{center}
\leavevmode
\epsfxsize=0.4\linewidth
\epsfbox{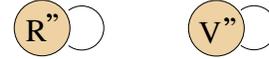}
\end{center}
\caption{Diagrams arising in first order of the cumulant
  expansion. The closed lines represent $G_>(\bN)$.  Both diagrams
  contribute to the free energy only, see Eq.  (\ref{1order}).}
\label{fig_diag1}
\end{figure}

After decomposition into slow and rapid fields, the average over the
fast modes in Eq. (\ref{fourier_pert}) can be easily performed using
Wick's theorem. For simplicity, we will drop the label $<$ on the slow
fields in the averaged expressions,
\begin{eqnarray}
\label{1order_av}
&-&\langle\cH_{\rm pert} \rangle_{>,c}\nonumber\\
&&=\frac{1}{2T^2}
\sum_{\alpha\beta}\int_{\bx,\bk} e^{i\bk
  \bphi_{\alpha\beta}(\bx)}
e^{-\bk^2TG_>(\bN)(1-\delta_{\alpha\beta})}\tR(\bk)\nonumber\\
&&+\frac{1}{T}\sum_\alpha \int_{\bx,\bk}e^{i\bk\bphi_\alpha(\bx)}
e^{-\frac{1}{2}\bk^2 T G_>(\bN)}\tV(\bk).
\end{eqnarray}
Here we have introduced the free two-point function
\begin{equation}
  \label{2point}
  G_>(\bx)=\int_\bq^> \frac{e^{i\bq\bx}}{\bq^2}\equiv
\int\limits_{|\bq|=\Lambda e^{-dl}}^\Lambda \frac{d^D \bq}{(2\pi)^D}
\frac{e^{i\bq\bx}}{\bq^2},
\end{equation}
where the label $>$ on the integral denotes integration over the
infinitesimal momentum shell. In the limit $dl\to 0$ this function, at
$\bx=\bN$, becomes
\begin{equation}
  G_>(\bN)=(2\pi)^{-D}S_D \Lambda^{D-2} dl\equiv K_D \Lambda^{D-2} dl,
\end{equation}
where $S_D$ is the surface area of a unit sphere in $D$ dimensions.
Since we have to calculate the expectation value of the perturbation
to first order in $dl$, the exponentials in Eq. (\ref{1order_av})
depending on the two-point function $G_>(\bN)$ can be expanded with
respect to $dl$.  This yields
\begin{eqnarray}
\label{1order}
  &-&\langle\cH_{\rm pert} \rangle_{>,c}\nonumber\\
&&=\frac{1}{2T^2}\sum_{\alpha\beta} \int_\bx
R(\bphi_{\alpha\beta})+\frac{1}{T}\sum_\alpha \int_\bx
V_L(\bphi_\alpha)\nonumber\\
&&+ \frac{K_D \Lambda^{D-2} dl}{2T} \int_\bx \left\{\sum_{\alpha\beta}
\partial_i\partial_i R(\bphi_{\alpha\beta})-n\partial_i\partial_i
R(\bN)\right\}\nonumber\\
&&+\frac{K_D \Lambda^{D-2} dl}{2} \int_\bx \sum_\alpha
\partial_i\partial_i V_L(\bphi_\alpha).
\end{eqnarray}
The corresponding diagrams are shown in Fig. \ref{fig_diag1}.  Here
and below the partial derivatives $\partial_i$ act on the field
$\bphi$, and repeated Latin indices are summed over from $1$ to $N$.
Whereas the first two terms are simply the original perturbation, the
remaining new parts are generated by the RG. The first and the last
term of this new part is of the appropriate form to renormalize
$R(\bphi)$ and $V_L(\bphi)$ but they are reduced by a factor of $T$
compared to the original ones. Thus they are irrelevant at the zero
temperature fixed points. The second new term proportional to $n$
contributes only to the renormalization of the free energy. Therefore
we conclude that in first order of the cumulant expansion there is no
renormalization of $R(\bphi)$ or $V_L(\bphi)$.

\begin{figure}[htb]
\begin{center}
\leavevmode
\epsfxsize=1.0\linewidth
\epsfbox{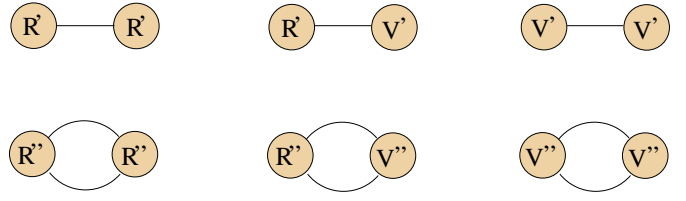}
\end{center}
\caption{Diagrams arising in second order of the cumulant
  expansion, where the internal lines represent $G_>(\bx-\bx')$.  The
  graphs with one internal line contribute only at large momenta and
  can be neglected.}
\label{fig_diag2}
\end{figure}

Now we turn to the analysis of the {\it second order} contributions to
the renormalized Hamiltonian. Averaging over the fields $\bphi^>$,
this contribution becomes
\begin{eqnarray}
  \label{2order_av}
&&\frac{1}{2}\langle\cH_{\rm pert}^2 \rangle_{>,c}\nonumber\\
&&=\int_{\bx_1,\bk_1 \atop \bx_2,\bk_2}\bigg\{
\frac{1}{8T^4}\sum_{\alpha_1 \beta_1 \atop \alpha_2 \beta_2}
e^{i\bk_1\bphi_{\alpha_1\beta_1}(\bx_1)+i\bk_2\bphi_{\alpha_2\beta_2}
(\bx_2)}\nonumber\\
&&\times e^{-\bk_1^2 TG_>(\bN)(1-\delta_{\alpha_1\beta_1})-\bk_2^2
TG_>(\bN)(1-\delta_{\alpha_2\beta_2})}\nonumber\\
&&\times e^{-\bk_1\bk_2 TG_>(\bx_1-\bx_2)(\delta_{\alpha_1\alpha_2}
+\delta_{\beta_2\beta_2}-\delta_{\alpha_1\beta_2}-\delta_{\beta_1\alpha_2}
)}\tR(\bk_1)\tR(\bk_2)\nonumber\\
&&+\frac{1}{2T^3}\sum_{\alpha_1\beta \atop\alpha_2} 
e^{i\bk_1\bphi_{\alpha_1\beta_1}(\bx_1)+i\bk_2\bphi_{\alpha_2}(\bx_2)}
e^{-\bk_1^2TG_>(\bN)(1-\delta_{\alpha_1\beta})}\nonumber\\
&&\times e^{-\frac{1}{2}\bk_2^2TG_>(\bN)}e^{-\bk_1\bk_2TG_>(\bx_1-\bx_2)
(\delta_{\alpha_1\alpha_2}-\delta_{\beta\alpha_2})}\tR(\bk_1)\tV(\bk_2)
\nonumber\\
&&+ \frac{1}{2T^2}\sum_{\alpha_1\alpha_2}
e^{i\bk_1\bphi_{\alpha_1}(\bx_1)+i\bk_2\bphi_{\alpha_2}(\bx_2)}
e^{-\frac{1}{2}(\bk_1^2+\bk_2^2)TG_>(\bN)}\nonumber\\
&&\times e^{-\bk_1\bk_2TG_>(\bx_1-\bx_2)\delta_{\alpha_1\alpha_2}}
\tV(\bk_1)\tV(\bk_2)\bigg\} - {\rm d.t.},
\end{eqnarray}
where d.t. stands for disconnected terms which have to be subtracted
from the above expression. To obtain the contributions to first order
in $dl$, we have to expand the exponentials depending on $G_>$ up to
second order in $T$. Sorting out the disconnected parts, the terms
proportional to $G_>(\bN)$ are canceled. Performing the summation over
the replica indices and transforming back to real space, we obtain the
following second order result,
\begin{eqnarray}
\label{2order_real}
  &&\frac{1}{2}\langle\cH_{\rm pert}^2 \rangle_{>,c}\nonumber\\
&&=\int_{\bx,\bx'}\bigg\{ \frac{1}{2T^3}\sum_{\alpha\beta\gamma}
\partial_iR(\bphi_{\alpha\beta}(\bx))\partial_i
R(\bphi_{\alpha\gamma}(\bx'))\nonumber\\
&&+\frac{1}{T^2}\sum_{\alpha\beta}\partial_i
R(\bphi_{\alpha\beta}(\bx))\partial_iV_L(\bphi_\alpha(\bx'))\nonumber\\
&&+\frac{1}{2T}\sum_\alpha \partial_i V_L(\bphi_\alpha(\bx))
\partial_i V_L(\bphi_\alpha(\bx'))\bigg\}G_>(\bx-\bx')\nonumber\\
&&+\int_{\bx,\bx'}\bigg\{ \frac{1}{4T^2}\sum_{\alpha\beta\gamma}
\partial_i\partial_jR(\bphi_{\alpha\beta}(\bx))
\partial_i\partial_jR(\bphi_{\alpha\gamma}(\bx'))\nonumber\\
&&+\frac{1}{4T^2}\sum_{\alpha\beta}\left[
\partial_i\partial_jR(\bphi_{\alpha\beta}(\bx))
\partial_i\partial_jR(\bphi_{\alpha\beta}(\bx'))\right.\nonumber\\
&&\left.-2\partial_i\partial_jR(\bphi_{\alpha\beta}(\bx))
\partial_i\partial_jR(\bN)\right]\nonumber\\
&&-\frac{1}{2T}\sum_\alpha
\partial_i\partial_jR(\bN)\partial_i\partial_j
V_L(\bphi_\alpha(\bx'))\nonumber\\
&&+\frac{1}{4}\sum_\alpha\partial_i\partial_j
V_L(\bphi_\alpha(\bx))\partial_i\partial_jV_L(\bphi_\alpha(\bx'))
\bigg\}G_>^2(\bx-\bx')\nonumber\\
\end{eqnarray}
corresponding to the diagrams in Fig. \ref{fig_diag2}.  The three
terms within the first curly bracket have a contribution only at large
momenta since $G_>(\bx)$ is composed only of momenta within the shell.
The first of these contributions is a 3-replica term which does not
fit into the form of the original replica-Hamiltonian (\ref{FH_n}).
Such terms with more than two replicas result from non-Gaussian
correlations in the disorder and are generated by the RG.
Anticipating that $R(\bphi)$ will be of order $\epsilon$ at the fixed
points, the renormalization of $R(\bphi)$ and $V_L(\bphi)$ resulting
from 3-replica parts will be of higher order in $\epsilon$ since the
3-replica terms are generated only at large momenta. Similarly, one
can argue that also the other two large momenta terms in the first
curly bracket do not influence the RG results to lowest order in the
small parameters of the RG expansion. Note that it may become
necessary to introduce a second small parameter which controls the
fixed point value of the lattice potential $V_L(\bphi)$. Therefore, we
will drop in the following all terms generated by the RG which
contribute only at large momenta.

From the terms in the second curly brackets, the first one which has
three replica indices would renormalize terms with three replicas
which would appear in the original Hamiltonian with a temperature
dependence $\sim T^{-3}$ such that the above term can be neglected at
a $T=0$ fixed point.  The next two terms are relevant and lead to a
renormalization of $R(\bphi)$ and $V_L(\bphi)$, respectively. The last
term is also of the right form to feed back into $V_L(\bphi)$, but it
is down by a factor of $T$ and hence negligible.

To evaluate the two relevant contributions of Eq. (\ref{2order_real}),
we note that main contribution to the $\bx'$-integration comes from
$\bx'=\bx$ since $G_>(\bx-\bx')$ then becomes maximal. Therefore, we
make of change of coordinates according to $\bx'\to \bx+\bx'$ and
expand the interactions with respect to $\bx'$. This gives formally
expressions of the form
\begin{eqnarray}
  \label{local_exp}
&&\int_{\bx,\bx'}\!\!\! F(\bx)F(\bx')G_>^2(\bx-\bx')=\int_{\bx,\bx'} 
\!\!\!F(\bx)F(\bx+\bx')G_>^2(\bx')\nonumber\\
&&=\int_\bx F^2(\bx) \int_{\bx'}
G_>^2(\bx')+\int_\bx F(\bx)\bbox{\nabla}F(\bx)\cdot \int_{\bx'} \bx'
G_>^2(\bx')\nonumber\\
&&+\frac{1}{2}F(\bx)\partial_{\bx_n}\partial_{\bx_m}F(\bx)\int_{\bx'}
\bx'_n \bx'_m G_>^2(\bx')+{\cal O}({\bx'}^3),
\end{eqnarray}
where $n$ and $m$ are summed over from $1$ to $D$.  The part which is
linear in $\bx'$ vanishes due to the symmetry of $G_>(\bx')$.  Since
$G_>(\bx)$ is composed only of momenta of the shell, the kernel
$G_>^2(\bx)$ can be easily integrated to first order in $dl$,
\begin{equation}
  \label{kernel_eval}
  \int_\bx G_>^2(\bx)=K_D\Lambda^{D-4}dl.
\end{equation}
Generally, one may expect also the appearance of quadratic gradient
terms from second derivatives of $F(\bx)$ in the above expansion,
resulting in a renormalization of the elastic stiffness constant
$\gamma$.  Whereas the absence of any $V_L$ terms in the relevant
contributions $\sim R^2$ renders gradient terms strongly irrelevant,
they have to be taken into account for the $RV_L$ part. Up to second
order in $\bbox{\nabla \phi}$, the non-vanishing parts of a gradient
expansion of the penultimate term of Eq. (\ref{2order_real}) are
\begin{eqnarray}
  \label{2order_grad_exp}
  &-&\frac{1}{2T}\sum_\alpha\int_{\bx,\bx'}\partial_i\partial_j R(\bN)
\Big\{\partial_i\partial_j V_L(\bphi_\alpha(\bx))\nonumber\\
&&+\frac{1}{2} \partial_i\partial_j \partial_k\partial_l 
V_L(\bphi_\alpha(\bx))\bx'\bbox{\nabla}\phi^k_\alpha(\bx)\cdot
\bx'\bbox{\nabla}\phi^l_\alpha(\bx)\Big\}G_>^2(\bx').\nonumber\\
\end{eqnarray}
The first term contributes to the renormalization of $V_L$ whereas
from the last part one may expect a change of the elastic stiffness.
Thus to lowest order in the gradient expansion, we end up with the
following contributions,
\begin{eqnarray}
  \label{2order_sum}
  \delta^{(2)}R&=&\frac{1}{2}K_D\Lambda^{D-4}dl\left\{\partial_i\partial_j
R(\bphi)\partial_i\partial_jR(\bphi)\right.\nonumber\\
&&\left.-2\partial_i\partial_jR(\bphi)
\partial_i\partial_jR(\bN)\right\}\nonumber\\
\delta^{(2)}V_L&=&-\frac{1}{2}K_D\Lambda^{D-4}dl\partial_i\partial_jR(\bN)
\partial_i\partial_j V_L(\bphi),
\end{eqnarray}
where $\delta^{(n)}R$ and $\delta^{(n)}V_L$ denote here and in the
following the $n^{\rm th}$ cumulant order feedback to the
renormalization of $R(\bphi)$ and $V_L(\bphi)$, respectively.

To analyze the last term in Eq. (\ref{2order_grad_exp}) we note that
$\partial_i\partial_jV_L(\bphi)\sim\delta_{ij}$ since we have assumed
the particularly simple form (\ref{defL}) for the lattice potential.
Therefore the diagonal structure of the elastic kernel is preserved.
To obtain the relevant contribution to the stiffness constant
$\gamma$, the gradient term has to be expanded in terms of
eigenfunctions of the linear operator of the RG flow of $V_L(\bphi)$
\cite{Balents+94}. This linear operator is given by
\begin{equation}
  \label{lin_V_op}
  {\cal L}V_L=(2-2\zeta)V_L+\zeta\phi^i\partial_iV_L-\frac{1}{2}
\partial_i\partial_jR(\bN)\partial_i\partial_jV_L,
\end{equation}
where the last term stems from the second order contribution
(\ref{2order_sum}) and the remaining parts arise from rescaling. The
determination of eigenfunctions is simple since
$\partial_i\partial_jR(\bN)=-\delta_{ij}\Delta$ in all cases of
interest. It is simple to see that the eigenfunctions are given by
\begin{equation}
  \label{V_eigenf}
  E_{\bf m}(\bphi)=\cos(pm_1\phi^1)\cdot\ldots\cdot\cos(pm_N\phi^N)
\end{equation}
with ${\bf m}=(m_1,\ldots,m_N)$ a vector of integers. Here we have
taken into account the rescaling of $\bphi(\bx)$ by allowing for a
periodicity $a$ of the lattice potential which depends on the flow
parameter $l$ such that $da/dl=-\zeta a$. Then the eigenfunctions
multiplied by the simple exponential factor $\exp(\lambda_{\bf m}l)$
are solutions of the linear RG flow $dV_L/dl={\cal L}V_L$. The scaling
dimension of this eigenfunctions at a fixed point is given by the
corresponding eigenvalues
\begin{equation}
  \label{V_eigenvals}
  \lambda_{\bf m}=2-2\zeta -\frac{1}{2}{\bf m}^2\lim_{l\to\infty}
p^2\Delta.
\end{equation}
It will turn out below that at these fixed points, where the lattice
potential $V_L(\bphi)$ has a finite value, the eigenvalues become
$\lambda_{\bf m}=2(1-{\bf m}^2)$ up to corrections of the order of the
small parameters of the RG expansion, e.g. $\epsilon$. Therefore, we
see that higher harmonics of the lattice potential with ${\bf m}^2>1$
are strongly irrelevant for the large scale behavior of the elastic
object. This justifies the assumption of Eq. (\ref{defL}) for the
lattice potential.

To analyze the last term of Eq. (\ref{2order_grad_exp}), this term has
to be written as a linear combination of the operators
\begin{equation}
\label{K_operators}
K_{\bf m}=\frac{1}{T} \int_\bx E_{\bf m}(\bphi)\bbox{\nabla \phi}\cdot
\bbox{\nabla \phi}.
\end{equation}
For ${\bf m}^2>0$, all these operators are strongly irrelevant with
scaling dimensions $\lambda_{\bf m}-2+2\zeta$ at the fixed points with
finite lattice potential. Therefore only the projection of the
gradient term in Eq. (\ref{2order_grad_exp}) onto the constant
eigenfunction $E_{\bf 0}(\bphi)$ can potentially contribute to the
renormalization of the stiffness constant. But this projection
vanishes since $V_L(\bphi)$ is simply a sum of first harmonics. Thus
$\gamma$ is not renormalized to first order in the lattice potential,
and we have no contribution to $\gamma$ from second cumulant order,
\begin{equation}
\delta^{(2)}\gamma=0
\end{equation}
where $\delta^{(n)}$ denotes the n$^{\rm th}$ order feedback to
$\gamma$.

To obtain a finite fixed point for $V_L(\bphi)$, we have to calculate
the RG flow up to the first non-vanishing non-linear order in
$V_L(\bphi)$. Therefore, we have to consider now contributions from
{\it third order} of $\cH_{\rm pert}$. In this order only terms of the
order $RV_L^2$ are relevant as will be explained now. The terms $\sim
R^3$ generate only three-replica parts which will be of order
$\epsilon^2$ and can therefore be neglected. On the other hand, the
contributions $\sim V_L^3$ which would be able to renormalize $V_L$
are reduced by a factor of $T$ and thus are also negligible at the
zero-temperature fixed point. For the terms of the order $R^2V_L$ the
discussion is slightly more complicated. The corresponding two-replica
terms generated in this order are of the form
\begin{equation}
  \label{irr_terms}
  \frac{1}{T^2}\int_\bx\sum_{\alpha\beta} \partial_i\partial_j 
R(\bphi_{\alpha\beta})\partial_i\partial_k 
R(\bphi_{\alpha\beta})\partial_j\partial_k V_L(\bphi_\alpha),
\end{equation}
which does not fit into the form of the terms present in the original
Hamiltonian. To analyze this unusual term, one has to expand again the
lattice potential $V_L(\bphi)$ into eigenfunctions of the linear
operator of the RG flow equation for $V_L(\bphi)$. This will enable us
to determine the relevance of contributions of the form
(\ref{irr_terms}). For that purpose we rewrite (\ref{irr_terms}) as a
linear combination of operators defined by
\begin{equation}
  S_{\bf m}=\frac{1}{T^2}\int_\bx \sum_{\alpha\beta} \partial_i\partial_j 
R(\bphi_{\alpha\beta})\partial_i\partial_k 
R(\bphi_{\alpha\beta})E_{\bf m}(\bphi_\alpha).
\end{equation}
The scaling dimension of these operators can easily be obtained in
terms of $\lambda_{\bf m}$. It turns out that the operators $S_{\bf
  m}$ for all ${\bf m}$ are strongly irrelevant with scaling dimensions
$-D-2-2\zeta+\lambda_{\bf m}=-4-2{\bf m}^2+{\cal O}(\epsilon)$.
Therefore, terms of the form (\ref{irr_terms}) are generated by the RG
but are negligible for the large-scale behavior.

\begin{figure}[htb]
\begin{center}
\leavevmode
\epsfxsize=1.0\linewidth
\epsfbox{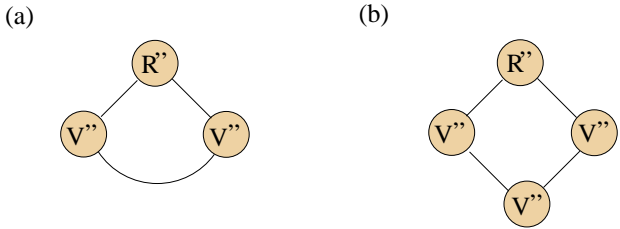}
\end{center}
\caption{Relevant graphs which appear in third (a) and fourth order
  (b) of the cumulant expansion. In both cases the disorder vertex
  contributes only a $\bphi$-independent factor
  $\partial_i\partial_jR(\bN)$.}
\label{fig_diag3}
\end{figure}

The only terms in third order remaining to be discussed are
proportional to $RV_L^2$. In contrast to the other combinations of $R$
and $V_L$ treated above, these terms produce a relevant feedback to
renormalize $V_L$. The corresponding part of $\langle\cH_{\rm pert}^3
\rangle_{>,c}$ is given by an expression similar to those appearing in
Eq. (\ref{2order_av}). In third order the relevant contributions come
from the $T^3$ terms of the expansion of the exponentials containing
the function $G_>(\bx)$. Integrating out the rapid modes and
transforming back to real space, we end up with the result
\begin{eqnarray}
  \label{3order_av}
&&\frac{1}{2T}\int_{{\bx_1\atop\bx_2}\atop\bx_3}
\sum_\alpha \partial_i\partial_jR(\bN)
\partial_i\partial_k V_L(\bphi(\bx_2))\partial_j\partial_k
V_L(\bphi(\bx_3))\nonumber\\
&& \times G_>(\bx_1-\bx_2)G_>(\bx_1-\bx_3)G_>(\bx_2-\bx_3)
\end{eqnarray}
represented by the graph in Fig. \ref{fig_diag3}(a).  We obtain in
lowest order of the gradient expansion in Eq.  (\ref{local_exp}) the
following third order contribution to $V_L(\bphi)$,
\begin{equation}
  \label{3order_sum}
  \delta^{(3)}V_L=-\frac{1}{2}K_D\Lambda^{D-6}dl\partial_i\partial_jR(\bN)
\partial_i\partial_k V_L(\bphi)\partial_j\partial_k V_L(\bphi).
\end{equation}
But for this sort of feedback also higher order terms of the gradient
expansion are relevant. The corresponding expression in second order
in $\bbox{\nabla \phi}$ is given by
\begin{eqnarray}
&&\frac{1}{2T}\!\!\int_{{\bx_1\atop\bx_2}\atop\bx_3}
\!\!\!\sum_\alpha \partial_i
\partial_j R(\bN)\partial_i\partial_k\partial_l
V_L(\bphi_\alpha(\bx_1))\partial_j\partial_k\partial_m
V_L(\bphi_\alpha(\bx_1)) \nonumber\\
&& \times \bx_2\bbox{\nabla}\phi^l_\alpha(\bx_1)\cdot
\bx_3\bbox{\nabla}\phi^m_\alpha(\bx_1)G_>(\bx_2)G_>(\bx_3)
G_>(\bx_2-\bx_3).\nonumber\\
\end{eqnarray}
As before, the diagonal structure of the elastic part is preserved by
this feedback due to $\partial_i\partial_j V_L\sim \delta_{ij}$. In
contrast to the second part of Eq. (\ref{2order_grad_exp}), this
contribution is quadratic in $V_L$. Therefore, also the first
harmonics of the lattice potential yield a non-vanishing projection
onto the constant eigenfunction $E_{\bf 0}(\bphi)$ since
$\sin^2(p\phi)=(1-\cos(2p\phi))/2$. Performing the spatial integration
over $\bx_2$ and $\bx_3$ to lowest order in $dl$ and $\epsilon$, we
obtain the relevant feedback to the stiffness constant from third
cumulant order,
\begin{eqnarray}
  \label{3order_gamma}
  \delta^{(3)}\gamma&=&-K_D\Lambda^{D-8}dl \partial_i\partial_j R(\bN)
\nonumber\\
&\times&{\cal P}_0[\partial_i\partial_k\partial_l
V_L(\bphi)\partial_j\partial_k\partial_l V_L(\bphi)].
\end{eqnarray}
Here ${\cal P}_0$ denotes the projection operator onto the
eigenfunction $E_{\bf 0}(\bphi)$.

Below we prefer to keep the stiffness constant fixed by allowing
the temperature to flow, so we have to rewrite the feedback
(\ref{3order_gamma}) as a renormalization of $T$. But this means that
also the feedback to the functions $R(\bphi)$ and $V_L(\bphi)$ is
changed due to the overall factor $1/T$ in front of the Hamiltonian.
The effective but non-rescaled quantities are then given to first
order in $dl$ by
\begin{eqnarray}
  \label{eff_T}T_{\rm eff}&=&(1-\delta^{(3)}\gamma)T,\\
  \label{eff_R}R_{\rm eff}&=&(1-2\delta^{(3)}\gamma)R+\delta^{(2)}R,\\
  \label{eff_V}V_{L,{\rm eff}}&=&(1-\delta^{(3)}\gamma)V_L+
  \delta^{(2)}V_L+\delta^{(3)}V_L.
\end{eqnarray}
The part $\delta^{(3)}\gamma$ is quadratic in $V_L$, and it is
tempting to neglect this term in Eq. (\ref{eff_V}) since it produces
cubic $V_L$ terms which appear irrelevant compared to the term
$\delta^{(3)}V_L$ which is quadratic in $V_L$. But it turns out that
this quadratic term does not contribute to the RG flow of the first
and only relevant harmonics of the lattice potential. The reason for
this is that a squared first harmonic produces only a constant and a
second order harmonic after projecting onto the eigenfunctions $E_{\bf
  m}(\bphi)$.

As explained above, the lowest non-linear $V_L$ term in the flow
equation of $V_L$ will be of third order. Since all relevant
contributions have to include at least one disorder vertex $\sim R$,
we have to consider also possible feedback to $V_L$ from the {\it
  fourth order} cumulant.  Simple counting of the $R$ and $V_L$ orders
suggests that in this order terms proportional to $R^2 V_L^2$ and $R
V_L^3$ have to be taken into account. For the first type of feedback
similar arguments as those for the contribution of Eq.
(\ref{irr_terms}) show that it is irrelevant in the RG sense.  In
contrast, the second kind of terms generates a relevant
renormalization of $V_L$. This feedback can again be calculated along
the lines described above. Now the relevant connected low-momenta
parts come from the $T^4$ terms of the expansion of the exponentials
with respect to the two-point function $G_>(\bx)$.  To first order in
$dl$ the $4^{\rm th}$ order contribution becomes
\begin{eqnarray}
  \label{4order_sum}
  \delta^{(4)}V_L&=&-\frac{1}{2}K_D\Lambda^{D-8}dl\partial_i\partial_j
  R(\bN)\partial_i\partial_k V_L(\bphi)
 \partial_j\partial_m V_L(\bphi)\nonumber\\
&\times& \partial_k\partial_m V_L(\bphi).
\end{eqnarray}
This part follows from the lowest order of the gradient expansion in
Eq. (\ref{local_exp}); higher gradient terms are irrelevant since they
could lead to a renormalization of $\gamma$ only of third order in
$V_L$ and, therefore, could produce higher order terms only in the
flow equation for $V_L$. Therefore contributions to $\gamma$ of higher
than second order in $V_L$ can be neglected.

Moreover, it is important to note that in higher than fourth order of
the cumulant expansion no relevant terms are generated which
renormalize $R$ and $V_L$ to lowest non-linear order. For example, in
fifth order terms appear which are cubic in $V_L$ and of the right
form to renormalize $V_L$ but with a coefficient proportional to
$R^2$. Therefore we can stop our analysis in fourth order of the
cumulant expansion.

Keeping track of all relevant feedbacks calculated above and
introducing the function
\begin{equation}
  \label{def_W}
  W(\bphi)=\partial_i\partial_j R(\bN) \partial_i\partial_k \partial_m
  V_L(\bphi)\partial_j\partial_k \partial_m V_L(\bphi),
\end{equation}
we obtain the resulting functional RG equations, valid to lowest order
in $\epsilon=4-D$,
\begin{eqnarray}
\label{frg_T}\frac{dT}{dl}&=&(2-D-2\zeta+{\cal P}_0[W(\bphi)])T\equiv
-\theta T,\\
\label{frg_R}
\frac{dR(\bphi)}{dl}&=&(4-D-4\zeta)R(\bphi)+\zeta
\phi^i \partial_i R(\bphi)\nonumber\\
+&&\!\!\!\!\!\!\!\frac{1}{2}\partial_i\partial_j
R(\bphi)\partial_i\partial_j R(\bphi)
-\partial_i\partial_j R({\bf 0})\partial_i\partial_j R(\bphi)\nonumber\\
+&&\!\!\!\!\!\!\! 2{\cal P}_0[W(\bphi)]R(\bphi),\label{Rflow}\\
\label{frg_V}
\frac{dV_L(\bphi)}{dl}&=&(2-2\zeta)V_L(\bphi)
+\zeta \phi^i \partial_i V_L(\bphi)\nonumber\\
-&&\!\!\!\!\!\!\!\frac{1}{2}\partial_i\partial_j R({\bf 0})\left[ 
\partial_i\partial_j V_L(\bphi)
+\partial_i\partial_k V_L(\bphi)\partial_j\partial_k
V_L(\bphi)\right. 
\nonumber\\ 
+&&\!\!\!\!\!\!\! \partial_i\partial_k V_L(\bphi)
 \partial_j\partial_m V_L(\bphi)
\partial_k\partial_m V_L(\bphi) ]\nonumber\\
+&&\!\!\!\!\!\!\! {\cal P}_0[W(\bphi)]V_L(\bphi),
\end{eqnarray}
where we made the replacements $\Lambda^{-2}V_L(\bphi) \rightarrow
V_L(\bphi)$, $K_D \Lambda^{D-4} R(\bphi) \rightarrow R(\bphi)$ with
$K_D^{-1}=2^{D-1}\pi^{D/2}\Gamma(D/2)$.

\section{Absence of lattice pinning: The rough phase}
\label{sec5}

Much is known about the equilibrium behavior of elastic manifolds and
periodic elastic media embedded in a disordered background for $D$
just below the upper critical dimension four. In the following, we
will briefly summarize the known results for the roughness exponent
$\zeta$ in the various cases. In addition, we will develop a
simplified calculation scheme to obtain the effective disorder
strength or, equivalently, the RG flow of the disorder potential on
sufficiently large but finite length scales. This technique becomes
useful if we study the competition between random and crystal pinning
in Section \ref{sec6}.

In the absence of lattice pinning, the RG flow of the disorder
correlator $R(\bphi)$ has been studied repeatedly in the past
\cite{FisherDS86b,Balents+93,Giamarchi+95}. The RG flow is determined
by Eq. (\ref{frg_R}) with $V_L(\bphi)\equiv 0$.  The fixed point
functions $R^*(\bphi)$ are always of the order $\epsilon$, but their
functional forms may be quite different depending on the bare
functions $R_0(\bphi)$ at $l=0$. There exist mainly three different
locally stable fixed points, each with its own basin of attraction.
The corresponding solutions of Eq.  (\ref{frg_R}) with $V_L(\bphi)=0$
will be analyzed in more detail for the different types of structure
and disorder in the following subsections. Within this analysis the
main focus is on the simplified scheme which allows for a reduction of
the functional flow equation to a flow of at most two parameters. This
novel representation will be helpful below for the inclusion of an
additional crystal potential in the RG scheme.

\subsection{Elastic Manifolds}

For the elastic manifold models, we have to distinguish between
long-range correlated and short-range correlated random potentials.
Both types of correlations can be physically realized for domain-walls
($N=1$) in, for example, disordered ferroelectrics or Ising magnets.
Whereas a random field leads to long-range correlated randomness,
dilution forms short range correlated random bond disorder.  In the
following, we will generalize our considerations to structures with
arbitrary $N$.  For the rest of this section we set $V_L(\bphi)=0$
which corresponds to $W(\bphi)=0$ in Eq.  (\ref{frg_R}).

\subsubsection{Random Fields}

It is easy to show that the $R(\bphi)\sim |\bphi|$ behavior for large
$|\bphi|$ of the bare disorder correlator is preserved under the RG
flow. Therefore we have to search for rotationally invariant solutions
of the flow Eq. (\ref{frg_R}) with this asymptotic behavior. Power
counting of this equation with respect to $\bphi$ gives then the
roughness exponent $\zeta^{\rm RF}=(4-D)/3$ independent of $N$. But we
are interested also in the fixed point function $R^*(\bphi)$ and the
RG flow toward it. Using the ansatz $R(\bphi)=\hat R(|\bphi|)$, the
flow equation reduces after $\hat R\to R$ to
\begin{eqnarray}
\label{reduced_R_flow}
\frac{dR}{dl}&=&(\epsilon-4\zeta)R(\phi)+\zeta\phi R'(\phi)+\frac{1}{2}
[R''(\phi)]^2 - R''(0)R''(\phi)\nonumber\\
&+& (N-1)\left\{\frac{1}{2}\frac{[R'(\phi)]^2}{\phi^2}
-R''(0)\frac{R'(\phi)}{\phi}\right\}.
\end{eqnarray}
The fixed point solution $R^*(\phi)$ of this equation can be
parameterized by its curvature $\Delta^* \equiv -R^{*''}(0)$, which is
a measure for the disorder strength, and by its characteristic length
scale $\xi^*$. With these parameters, the fixed point function can be
written as
\begin{equation}
\label{fp_ansatz}
R^*(\phi)=\Delta^*\xi^{*2}r(\phi/\xi^*).
\end{equation}
This relation defines the dimensionless function $r(u)$ with
$u=\phi/\xi^*$ and $r''(0)=-1$. It describes the {\it functional form}
of the fixed point solution and has a characteristic scale of order
unity. Notice that there is a hole family of fixed points which are
parameterized by different $\xi^*$ since one has the freedom to choose
an arbitrary overall length scale for $\bphi$ in the Hamiltonian
(\ref{genHam}) without crystal potential $V_L(\bphi)$. As we will see
below, once a particular value for $\xi^*$ has been chosen, the
corresponding $\Delta^*$ is automatically fixed. From Eqs.
(\ref{reduced_R_flow}) and (\ref{fp_ansatz}) follows that the function
$r(u)$ has to fulfill the equation
\begin{eqnarray}
\label{fp_form}
0&=&\alpha r(u) +\beta u r'(u) +\frac{1}{2}
[r''(u)]^2 + r''(u)\nonumber\\
&+& (N-1)\left\{\frac{1}{2}\frac{[r'(u)]^2}{u^2}
+\frac{r'(u)}{u}\right\}
\end{eqnarray}
where $\alpha=(\epsilon-4\zeta)\xi^{*2}/\Delta^*$ and
$\beta=\zeta\xi^{*2}/\Delta^*$ are numerical coefficients. From this
one obtains the exponent 
\begin{equation}
\zeta=\frac{\epsilon}{4+\alpha/\beta}.
\end{equation}
Generally, these coefficients are fixed first by the condition that
$r(u)$ has the correct asymptotic behavior and second by the choice of
an overall length scale for $\bphi$. From $r(u)\sim -|u|$ for large
$u$ follows the relation $\alpha=-\beta$. We choose a scale for
$\bphi$ such that $r(0)=-1$.  That $r(0)$ has to be negative can be
seen as follows.  Evaluating Eq. (\ref{fp_form}) for small $u$, one
easily obtains $\alpha=N/2r(0)$ due to $r''(0)=-1$. Since
$\beta=\zeta\xi^{*2}/\Delta^*>0$ we have $\alpha<0$ and our choice of
scale corresponds to $\alpha=-N/2$.

After we have determined the coefficients $\alpha$ and $\beta$, the
fixed point function can be calculated in principle at least
numerically by solving Eq. (\ref{fp_form}) for $r(u)$ with initial
conditions $r(0)=-1$ and $r'(0)=0$. But from a physical point of view,
we are also interested in the evolution of the disorder under the RG
transformation sufficiently close to the fixed point to study its
stability or to calculate critical exponents of an unstable fixed
point. In particular, this becomes important if we take into account a
competing crystal potential which may lead to a new unstable fixed
point associated with a roughening transition. Now, there is a {\it
  special} set of bare correlators $R_0(\phi)$ for which the RG flow
can be calculated on all length scales exactly. This special set is
given by the two parameter family of functions
$R_0(\phi)=\Delta_0\xi_0^2r(\phi/\xi_0)$, i.e., the functions are
already of the functional fixed point form but differ by their bare
strength $\Delta_0$ and bare characteristic width $\xi_0$. The RG flow
in the hyper-plane which is formed by these two parameters can be
obtained by inserting the Ansatz
\begin{equation}
  \label{R_ansatz}
  R(\phi)=\Delta(l)\xi^2(l)r(\phi/\xi(l))
\end{equation}
into the flow equation (\ref{reduced_R_flow}). Using the fact that
$r(u)$ satisfies Eq. (\ref{fp_form}), this results in the following
differential equation for $r(u)$,
\begin{equation}
  \label{diff_eq_r}
A(l)r(u)+B(l)ur'(u)=0,
\end{equation}
where
\begin{eqnarray}
  \label{AB_eq}
  A(l)&=&2\xi \Delta \frac{d\xi}{dl} + \xi^2 \frac{d\Delta}{dl}
  -(\epsilon-4\zeta)\xi^2\Delta+\alpha \Delta^2,\\
  B(l)&=&-\xi\Delta \frac{d\xi}{dl}-\zeta\xi^2\Delta+
  \beta\Delta^2.
\end{eqnarray}
In the last formulas we have dropped the $l$ dependence of $\xi$ and
$\Delta$. All the solutions $r(u)$ of Eq. (\ref{diff_eq_r}) are simple
powers of $u$ and vanish for $u=0$. But this is in contradiction to
the correct form of $r(u)$ determined by Eq. (\ref{fp_form}), which is
not a simple power of $u$ and moreover finite for $u=0$. Therefore
such power law solutions cannot occur, and Eq. (\ref{diff_eq_r}) is
fulfilled if and only if $A(l)=B(l)=0$ for all $l$. This condition
straightforwardly leads to the RG flow equations
\begin{eqnarray}
  \frac{d\Delta}{dl}&=&(\epsilon-2\zeta)\Delta-\frac{\alpha+2\beta}
        {\xi^2}\Delta^2,
  \label{flow_Delta_pure}\\
  \frac{d\xi}{dl}&=&\left(-\zeta+\beta\frac{\Delta}{\xi^2}\right)\xi.
  \label{flow_xi_pure}
\end{eqnarray}
Flow equations of the same form but with different coefficients have
been derived by Nattermann and Leschhorn \cite{Nattermann+91} for
random bond disorder, using the approximation that the bare Gaussian
shape of $R(\bphi)$ is preserved under the RG flow.

Before we come to the fixed point analysis of these equations, we will
comment on the advantage of the two parameter representation over the
functional flow of Eq. (\ref{reduced_R_flow}). For the Hamiltonian
(\ref{genHam}), without any perturbation from the crystal lattice, it
has been shown that there exists a family of random field fixed points
which is at least locally stable and represents the rough phase of the
interface \cite{FisherDS86b,Balents+93}. Since this family of fixed
points lies in the hyper-plane of the two parameter flow, the RG flow
starting from an arbitrary bare random field disorder correlator
$R_0(\phi)$ should be described on sufficiently large length scales
also by Eqs.  (\ref{flow_Delta_pure}), (\ref{flow_xi_pure}).

\begin{figure}[htb]
\begin{center}
\leavevmode
\epsfxsize=0.7\linewidth
\epsfbox{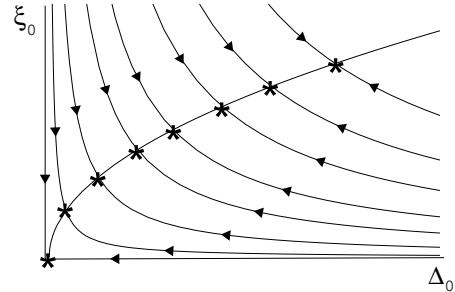}
\end{center}
\caption{Schematic RG flow to a line of fixed points in the 
hyper-plane formed by $\Delta_0$ and $\xi_0$.}
\label{fig_fp_line}
\end{figure}

Let us now turn back to the fixed point analysis of Eqs.
(\ref{flow_Delta_pure}), (\ref{flow_xi_pure}). As explained above, we
expect a line of stable fixed points resulting from the choice of the
overall length scale for $\phi$. This line is given by
\begin{equation}
  \label{fp_line}
  \Delta^*=\frac{\epsilon-2\zeta}{\alpha+2\beta} \xi^{*2},
\end{equation}
where
\begin{equation}
  \xi^*=\xi_0\left(\frac{\alpha+4\beta}{\epsilon}\frac{\Delta_0}
    {\xi_0^2}\right)^{\zeta/\epsilon}
\end{equation}
depends on the bare parameters $\xi_0$ and $\Delta_0$. The
corresponding RG flow is shown in Fig. \ref{fig_fp_line}. Linearizing
about the fixed points, one finds the universal eigenvalues
$\lambda_1=0$ and $\lambda_2=-\epsilon$ corresponding to a redundant
marginal and a stable direction, respectively. The leading irrelevant
eigenvalue of the full functional RG flow for random field disorder is
indeed given by $\lambda_2$ \cite{FisherDS86b}. But notice that in
general one cannot expect to obtain by the two parameter approach the
{\it leading} irrelevant eigenvalue out of the infinite number of
irrelevant eigenvalues.

The transversal displacement correlation function 
\begin{equation}
C({\bf x})\equiv \frac{1}{N}\overline{\langle 
[\bphi({\bf x})-\bphi({\bf 0})]^2 
\rangle}
\end{equation}
can be calculated perturbatively from the renormalized but
non-rescaled effective parameter
$\tilde\Delta(l)=\Delta(l)e^{(2\zeta-\epsilon)l}$ using
$l=\ln(\Lambda/q)$. On length scales larger than the Larkin length,
$|{\bx}|\gg
L_\Delta=[\epsilon N/(\alpha+4\beta)\hat\xi_0^2/\Delta_0]^{1/\epsilon}$, we
obtain
\begin{equation}
\label{rf_corr}
  C({\bf x})=(N/\beta)\hat\xi_0^2 (|{\bx}|/L_\Delta)^{2\zeta},
\end{equation}
where we have explicitly considered the $N$-depence of $\xi_0$ by
writing $\xi_0^2=N\hat\xi_0^2$.  Notice that the Larkin length
$L_\Delta$ and the coefficient of $C(\bx)$ remain finite for $N \to
\infty$ due to $\alpha=-\beta=-N/2$.

For $D=4$, the roughness exponent $\zeta$ vanishes and logarithmic
scaling of the transversal displacement is expected. From the RG flow
of the effective parameter $\tilde\Delta$ for $\epsilon=0$, we find
that the roughness grows only {\it sub-logarithmically},
\begin{equation}
\label{sub_ln}
C({\bf x})\sim \ln^\sigma(|\bx|/a_0), \quad 
\sigma=\frac{2\beta}{\alpha+4\beta}.
\end{equation}
For random field disorder, we have the exponent $\sigma=2/3$
independent of $N$.

\subsubsection{Random Bonds}

As for random field disorder, it is straightforward to see that the
functional behavior of the exponentially decaying correlator
$R_0(\bphi)$ for random bond disorder is preserved by the RG flow
\cite{Balents+93}. To obtain a valid fixed-point function for
short-range correlated disorder, therefore, we have to search for
solutions of Eq.  (\ref{reduced_R_flow}) with an exponential tail at
large $\phi$. The fixed-point function can still be written in the
form of Eq.  (\ref{fp_ansatz}), and it remains the problem to determine
the coefficients $\alpha$ and $\beta$ such that Eq. (\ref{fp_form})
has an exponentially decaying solution $r(u)$ for the initial
conditions $r(0)=1$, $r'(0)=0$. As in the random field case we have
$\alpha=N/2r(0)=N/2$ by our choice of scale.  Unfortunately, the
conditions for the behavior of $r(u)$ do not fix the coefficient
$\beta$ uniquely since there exists a discrete family of solutions of
Eq.  (\ref{fp_form}) which decay exponentially at infinity. But it can
be shown that only the function of the whole set of solutions
corresponding to the largest value of $\beta$ is stable with respect
to short-range perturbations \cite{Balents+93}. The corresponding
value for $\beta$ has to be calculated numerically.  Solving Eq.
(\ref{fp_form}) numerically, the unstable solutions can be
distinguished from the stable one since they show oscillations instead
of a simple exponential decay. The resulting values of $\beta$ are
shown in Tab. \ref{tab_beta} together with the corresponding roughness
exponents $\zeta$ for different $N$. The values for $\beta$ decrease
with $N$ and approach $1/2$ for $N\to\infty$ corresponding to
$\zeta/\epsilon \to 1/(4+N)$.

\begin{table}[h]
\begin{center}
\begin{tabular}{ccc}
\hline
$N$ & $\beta$ & $\zeta/\epsilon=(4+\alpha/\beta)^{-1}$ \\
\hline
1 & 0.6244 & 0.2083\\
2 & 0.6010 & 0.1766\\
3 & 0.5807 & 0.1519\\
4 & 0.5635 & 0.1325\\
5 & 0.5491 & 0.1169\\
\hline
\end{tabular}
\end{center}
\caption{\label{tab_beta}Numerical values for $\beta$ and the exponent
  $\zeta$.}
\end{table}

With these values for the coefficients $\alpha$ and $\beta$ the flow
equations Eqs. (\ref{flow_Delta_pure}) and (\ref{flow_xi_pure}) remain
valid also for the random bond case.  The RG flow is still given by
Eq.  (\ref{fp_line}) but with the exponent $\zeta$ from Tab.
\ref{tab_beta}, and looks similar to that shown in Fig.
\ref{fig_fp_line}. For $N<\infty$, the correlation function $C(\bx)$
is given by Eq. (\ref{rf_corr}) but with the values for $\zeta$ from
Tab. \ref{tab_beta}. In contrast, for $N=\infty$ the roughness
exponent vanishes and the manifold is only logarithmically rough,
\begin{equation}
  \label{rb_ln_corr}
   C(\bx)=4\epsilon \hat\xi_0^2 \ln(|\bx|/L_\Delta).
\end{equation}

The exponent $\sigma$ describing the
sub-logarithmic scaling of Eq. (\ref{sub_ln}) valid for $D=4$ becomes
$\sigma=0.4166$ for $N=1$ and approaches $\sigma=2/(N+4)$ for
$N\to\infty$.

\subsection{Periodic Elastic Media}

The periodicity of the unrenormalized $R_0(\bphi)$ has to be preserved
under the RG flow, hence we have to search for fixed point
solutions of Eq.  (\ref{frg_R}) with an appropriate symmetry given by
the reciprocal lattice vectors $\bQ$. In general, one has to treat
different lattice symmetries separately. For the simplest case, a
square lattice, the fixed point solution is separable,
\begin{equation}
  R^*(\bphi)=\sum_{n=1}^N \hat R^*(\phi^n),
\end{equation}
where $\hat R^*(\phi)$ is a periodic fixed point function of Eq.
(\ref{frg_R}) with $N=1$ and the $\phi^n$ are the components of
$\bphi$. For a triangular lattice the situation is more difficult and
we did not try to find a solution for this case.  The $N=1$ fixed
point solution $\hat R(\phi)$ corresponds to the charge density wave
case and has been calculated in Ref.  \cite{Giamarchi+95}. The RG flow
towards this fixed point can be analyzed by keeping track of the flow
of the Fourier coefficients of $\hat R(\phi)$.  For simplicity, we
choose the periodicity as $2\pi$ and take $\hat R(\phi) \to R(\phi)$.
Inserting the Fourier expansion
\begin{equation}
  \label{R_Fourier}
  R(\phi)=\sum_{m=1}^\infty R_m \cos(m\phi)
\end{equation}
into the functional flow Eq. (\ref{frg_R}) where we have to choose
$\zeta=0$ due to the periodicity, we obtain the following infinitely
many flow equations for the coefficients $R_m$ which are coupled to
each other,
\begin{eqnarray}
\frac{dR_m}{dl}&=&\epsilon R_m + \frac{1}{4}\sum_{m'=1}^{m-1}
m'^2(m-m')^2R_{m'}R_{m-m'}\nonumber\\
&+&\frac{1}{2}\!\!\sum_{m'=m+1}^{\infty}
\!\!\!\!m'^2(m'-m)^2 R_{m'}R_{m'-m}\nonumber\\
&-&m^2 R_m \sum_{m'=1}^{\infty} m'^2 R_{m'}.
\label{fourier_flow}
\end{eqnarray}
By performing simply the sums, it can be checked that a fixed point of
this set of equations is given by $R_m^*=2\epsilon/3m^4$. The
corresponding series (\ref{R_Fourier}) can be summed up and yields the
solution known from Ref. \cite{Giamarchi+95}. The flow towards
this fixed point as well as its stability can be analyzed by rewriting
the Fourier coefficients as
\begin{equation}
  \label{fourier_ansatz}
  R_m=\frac{6}{\pi^2}\frac{\Delta}{m^4}+r_m
\end{equation}
where $\Delta=-R''(0)$ with the fixed point value
$\Delta^*=\epsilon\pi^2/9$. Substituting this Ansatz into Eq.
(\ref{fourier_flow}), we see that the $r_m$ flow to zero with
eigenvalues $\lambda_m=\epsilon(2/3-m-2m^2/3)$ whereas the flow of
$\Delta$ is determined by
\begin{equation}
  \label{Delta_flow}
  \frac{d\Delta}{dl}=\epsilon\Delta-\frac{9}{\pi^2}\Delta^2.
\end{equation}
Since we are interested in the RG flow on sufficiently large length
scales, we will therefore assume here the particularly simple
functional form of the fixed point by dropping the $r_m$. This
approximation is justified on length scales larger than the
Fukuyama-Lee length $L_\Delta=\Delta^{-1/\epsilon}$ where the
parameters $r_m$ approach zero. The RG flow towards the fixed point is
then described by the single parameter flow of
Eq. (\ref{Delta_flow}).  

The transversal displacement correlation function growths only
logarithmically on scales beyond the Fukuyama-Lee length $L_\Delta$,
\begin{equation}
\label{corr_pm}
  C({\bf x})=2\frac{\pi^2}{9}\epsilon \ln(|\bx|/L_\Delta)
\end{equation}
with a universal coefficient which may depend on the actual lattice
symmetry.  The coefficient given here applies to a square lattice with
a periodicity of $2\pi$.

\section{The roughening transition}
\label{sec6}

To study the effect of a finite lattice pinning potential $V_L(\bphi)$
upon the large scale behavior of elastic objects, we have to
analyze the full set of functional RG equations
(\ref{frg_T})-(\ref{frg_V}). As has been shown is Section 5, while
the same functional RG equations apply for the different types of
elastic objects and bare disorder correlations, the corresponding
solutions are quite different. Therefore, we will discuss the analysis
of the RG flow in the following subsections separately for elastic
manifolds and elastic periodic media.

\subsection{Elastic manifolds}

For the case of elastic manifolds, the functional RG flow of the
disorder correlator $R(\bphi)$ has been reduced already to a flow of
two parameters for vanishing lattice pinning in Section 5. To develop
the RG analysis for finite $V_L(\bphi)$, it is important to note that
the functional form of the flow equation for $R(\bphi)$ is not
changed.  Indeed, a finite lattice potential leads only to a shift of
$\epsilon=4-D$ by the constant $2{\cal P}_0[W(\bphi)]$, as can be seen
easily from Eq. (\ref{frg_R}). Therefore, the RG flow of $R(\bphi)$
can be replaced again by the two-parameter flow of Eqs.
(\ref{flow_Delta_pure}) and (\ref{flow_xi_pure}) but with a shifted
$\epsilon$.  For the lattice potential we have to consider only the
first harmonics, and we assume in the following the simple form of Eq.
(\ref{defL}) with
\begin{equation}
\label{defV}
V(\phi)=v\cos\left(\frac{2\pi}{a}\phi\right)
\end{equation}
for all components of the vector $\bphi$.

To begin with, we calculate the projection of the function $W(\bphi)$
defined in Eq. (\ref{def_W}) onto the lowest harmonic
$E_\bN(\bphi)=1$.  Using Eq.  (\ref{defV}) and
$\partial_i\partial_jR(\bN)=-\Delta \delta_{ij}$, the function
$W(\bphi)$ becomes
\begin{equation}
  \label{res_W}
  W(\bphi)=-\Delta v^2 \frac{(2\pi)^6}{a^6} \sum_{i=1}^N \sin^2
  \left(\frac{2\pi}{a}\phi^i\right).
\end{equation}
After projection onto the lowest harmonic, this equation reduces to
\begin{equation}
  \label{proj_W}
  {\cal P}_0[W(\bphi)]=-\frac{1}{2}\Delta v^2 \frac{(2\pi)^6}{a^6}
\end{equation}
since $\sin^2(y)=(1-\cos(2y))/2$. Note that the negative eigenvalue
$\theta$ of the temperature flow, see Eq. (\ref{frg_T}), is also
shifted by this constant to a larger value. Physically, this
corresponds to an increased elastic stiffness of the manifold due to
the lattice potential.

The RG flow of the first harmonic of the lattice potential can be
obtained by inserting the sum of the first harmonics (\ref{defV}) for
$V_L(\bphi)$ in Eq. (\ref{frg_V}) and projecting again the whole
equation onto the first harmonics. To treat the second term $\sim
\phi^i$ in Eq. (\ref{frg_V}) properly, one has to rescale also the
periodicity $a$ which leads to the trivial flow $da/dl=-\zeta a$.
After the projection, the term which is quadratic in $V_L$ vanishes
since it contributes only to the second harmonics. Both the term cubic
in $V_L$ and the term ${\cal P}_0[W]V_L$ from the stiffness
renormalization are of order $v^3$. Using for the projection of $W$
the relation $\cos^3(y)=(3\cos(y)+\cos(3y))/4$, we obtain the
following parameter RG flow,
\begin{eqnarray}
  \label{rg1_T}
\frac{dT}{dl}&=&\left[2-D-2\zeta-\frac{1}{2}\Delta v^2\frac{(2\pi)^6}{a^6}
\right]T,\\
\label{rg1_D}
  \frac{d\Delta}{dl}&=&(\epsilon-2\zeta)\Delta-\left[
\frac{\alpha+2\beta}{\xi^2}+\frac{(2\pi)^6}{a^6}v^2\right]\Delta^2,\\
  \label{rg1_xi}
  \frac{d\xi}{dl}&=&\left[-\zeta+\beta\frac{\Delta}{\xi^2}\right]\xi,\\
\label{rg1_v}
\frac{dv}{dl}&=&\left[2-2\zeta-\frac{1}{2}\frac{(2\pi)^2}{a^2}\Delta
\right]v-\frac{7}{8}\frac{(2\pi)^6}{a^6}\Delta v^3,\\
\label{rg1_a}
\frac{da}{dl}&=&-\zeta a.
\end{eqnarray}
It is important to note that these flow equations are symmetric with
respect to $v \to -v$ as could be expected for a periodic potential
since this transformation corresponds only to a shift in $\phi^i$ by
$a/2$.  Naively, one expects for the coefficients $\alpha$ and $\beta$
the same relations as given below Eq. (\ref{fp_form}) but with shifted
$\epsilon$. But these relations have to be reexamined later since it
turns out that $\xi^{*2}/\Delta^*$ diverges at the new fixed point
describing the roughening transition.

To analyze the above scaling transformations, it is useful to
introduce the new variables $\bar\Delta\equiv (2\pi/a)^2\Delta$,
$\bar\xi\equiv (2\pi/a)\xi$ and $\bar v\equiv (2\pi/a)^2 v$. By this
change of variables, the roughness exponent $\zeta$ is eliminated from
the coupled flow equations.
To finally obtain $\zeta$, we have to evaluate the relations for
$\alpha$ or $\beta$ in terms of the flow of the new variables along
the critical line flowing into the fixed point. In terms of the new
variables, we have
\begin{eqnarray}
  \label{rg2_T}
  \frac{dT}{dl}&=&\left[2-D-2\zeta-\frac{1}{2}\bar\Delta \bar v^2\right]T,\\
\label{rg2_D}
  \frac{d\bar\Delta}{dl}&=&\epsilon\bar\Delta-\left[
\frac{\alpha+2\beta}{\bar\xi^2}+\bar v^2\right]\bar\Delta^2,\\
  \label{rg2_xi}
  \frac{d\bar\xi^2}{dl}&=&2\beta\bar\Delta,\\
\label{rg2_v}
\frac{d\bar v}{dl}&=&\left[2-\frac{1}{2}\bar\Delta
\right]\bar v-\frac{7}{8}\bar\Delta \bar v^3.
\end{eqnarray}

\begin{figure}[htb]
\begin{center}
\leavevmode
\epsfxsize=1.0\linewidth
\epsfbox{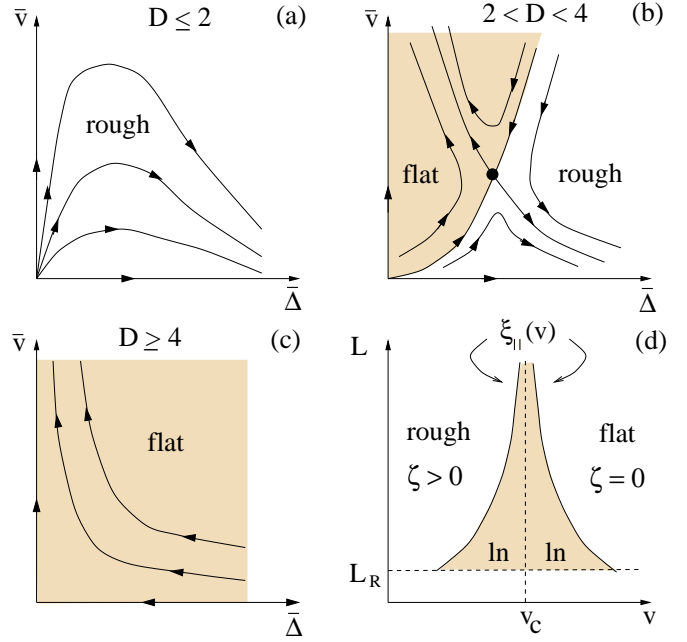}
\end{center}
\caption{Schematic RG flow for $N<\infty$ as a function of the
  disorder variable $\bar\Delta$ and the lattice strength $\bar v$ for
  (a) $D\le 2$, (b) $2<D<4$ and (c) $D\ge4$. Note that the fixed point
  associated with the rough phase is shifted here to
  $\bar\Delta=\infty$. (d) shows the length scale $L$ dependent
  roughness near the roughening transition where the correlation
  length $\xi_\|(v)$ diverges.}
\label{fig.mani_flow}
\end{figure}

Before we study the case $v\neq 0$, it should be mentioned that the
stable fixed point of the rough phase is located at $\bar v=0$, $\bar
\Delta=\infty$ in terms of the new variables since the periodicity $a$
flows exponentially fast to zero. Now we turn to the fixed point
analysis for finite $\bar v$ and $N<\infty$. The case $N=\infty$ will
be discussed separately in Section 6.4. To obtain a fixed point with a
finite value for $\bar v$ from Eq.  (\ref{rg2_v}), the parameter
$\bar\Delta$ has to flow along a critical line to a finite value
$\bar\Delta^*$. The corresponding fixed point value for $\bar v$ is
then given by $\bar v^{*2}=4(4/\bar\Delta^*-1)/7$. As a consequence of
Eq.  (\ref{rg2_xi}) and since $\beta>0$, one obtains the asymptotic
behavior $\bar\xi^2=2\beta\bar\Delta^*l$ up to sub-linear corrections.
But this divergence of $\bar\xi^2$ means that we can drop the first
term in the brackets of Eq. (\ref{rg2_D}) for the fixed point
analysis. This leads to the fixed point values, valid to lowest order
in $\epsilon$,
\begin{equation}
\label{fp_values}
\bar\Delta^*=4-\frac{7}{4}\epsilon, \quad \bar
v^{*2}=\frac{\epsilon}{4}.
\end{equation}
As one may expect, $\bar v^*$ vanishes when the upper critical
dimension is approached, justifying the $\epsilon$-expansion. It
should be noted that $\bar v^*$ has the meaning of a (rescaled) mass
for the field $\bphi$. The second small parameter in the double
expansion with respect to both pinning potentials is $\Delta$, which
is indeed small at the new fixed point since $\Delta\sim 1/l$ along
the critical line as we will see below. Moreover, it should be
mentioned that the fixed point value $\bar\Delta^*$ vanishes for
$\epsilon=16/7$ corresponding to a shift of this fixed point to the
$\bar v$ axis for $D=12/7$, see Fig. \ref{fig.mani_flow}. Note that
this is, despite the lowest order $\epsilon$ result, in approximate
agreement with the absence of a roughening transition for $D\le 2$ as
predicted by scaling arguments in Section 3.

Linearizing around this fixed point, we obtain, in lowest order in
$\epsilon$, the universal eigenvalues $\lambda_{\pm}=\pm
2\sqrt{\epsilon}$. Since the fixed point has an unstable direction, it
has to be associated with the roughening transition.  In the flat
phase, the size of typical excursions of the manifold from the
preferred minimum of the lattice potential defines the longitudinal
correlation length
\begin{equation}
\xi_\|\sim |v-v_{0,c}|^{-\nu_\|}, \quad \nu_\|=\frac{1}{2\sqrt{\epsilon}}.  
\end{equation}
Interestingly, $\nu_\|$ does not dependent on $\alpha$, $\beta$ and is
therefore universal for random field and random bond disorder. The
same expression for $\nu_\|$ has been obtained previously for the {\it
  thermal} roughening transition but with $\epsilon=2-D$
\cite{Forgacs+91}.

To determine the critical roughness exponent $\zeta_c$ at the
roughening transition, we have to re-examine the relations between the
coefficients $\alpha$ and $\beta$ and the fixed point values. Paying
attention to the fact that $\Delta$ approaches zero at the new fixed
point, we obtain the refined expression
\begin{equation}
  \label{coeff_alpha}
  \alpha=\lim_{l\to\infty} \frac{\bar\xi^2}{\bar\Delta}\left[
\epsilon-2\zeta-\bar\Delta\bar v^2 -\frac{1}{\bar\Delta}
\frac{d\bar\Delta}{dl}\right],
\end{equation}
which turns out to be equivalent to 
\begin{equation}
  \label{coeff_beta}
  \beta=\lim_{l\to\infty} \zeta\frac{\bar\xi^2}{\bar\Delta}.
\end{equation}
From the last expression we see that $\zeta$ has to flow along the
critical line asymptotically according to 
\begin{equation}
\zeta=\frac{1}{2l}
\end{equation}
since $\bar\xi^2/\bar\Delta=2\beta l$. Therefore, at the transition we
have $\zeta_c=0$ and the roughness of the manifold grows only {\it
  logarithmically}.  Due to $\zeta=1/2l$ for large $l$, the
periodicity of the lattice potential flows only algebraically to zero,
$a\sim l^{-1/2}$, in agreement with the logarithmic scaling at the
transition.  Therefore, the renormalized width $\xi\sim \bar\xi/a$ of
the disorder correlator flows to a finite value $\xi^*$ at the new fixed
point.  Notice that at this fixed point one has no longer the freedom
to choose an arbitrary overall length scale for $\bphi$ due to the
fixed periodicity $a_0$ of the lattice potential. As a consequence,
there is no line of fixed points connected by a redundant operator as
in the case of vanishing lattice potential.

The free energy fluctuations grow with length as $L^\theta$, where
$\theta$ is defined by Eq. (\ref{frg_T}). In the rough phase the tilt
symmetry of the Hamiltonian (\ref{genHam}) guarantees that the
temperature is renormalized only by scale changes. This leads to the
exact result $\theta=D-2+2\zeta$. The relevance of the lattice
potential results in an additional renormalization of temperature and
therefore a different exponent $\theta_c$ at the transition.
Independent of $\alpha$ and $\beta$, we obtain from the fixed point
values of Eq. (\ref{fp_values}) to lowest order in $\epsilon$ the
result
\begin{equation}
  \label{theta_c}
  \theta_c=2-\frac{\epsilon}{2}.
\end{equation}

The cross-over length scale $L_R$ introduced in Section 3 can also be
obtained from the RG flow of $\bar\Delta$. This length corresponds to
the scale where the roughness of the manifold changes from algebraic
to logarithmic growth on both sides of the transition. In terms of the
RG flow, $L_R=a_0e^{l_R}$ is determined by the scale $l_R$ where
$\bar\Delta(l)$ approaches its asymptotic fixed point value
$\bar\Delta^*$. To estimate $l_R$, one can integrate Eq.
(\ref{rg2_D}) with $\bar v=0$ to obtain a primary solution. The effect
of the $\bar v^2$-term can subsequently be included by matching the
primary solution $\bar\Delta(l)$ and $\bar\Delta^*$ at $l_R$. Doing
so, we get
\begin{equation}
  \label{L_R_result}
  L_R=L_\Delta\left(\frac{\alpha+4\beta}{N}\frac{\bar\Delta^*}{\epsilon}
\frac{a_0^2}{\hat\xi_0^2}\right)^{1/2\zeta},
\end{equation}
where $\zeta$ is the roughness exponent of the algebraic regime at
smaller scales.  It is given by the exponent $\zeta$ of the rough
phase for $\hat\xi_0\ll a_0$ and $\zeta=\epsilon/2$ for $\hat\xi_0\gg
a_0$.  This is in agreement with the scaling result of Eq.
(\ref{minL_R1}) if $v_0$ is replaced there by $v_{0,c}$ of Eq.
(\ref{v0}).

Beyond the length $L_R$ and sufficiently close to the transition where
$\xi_\| > |{\bf x}|$, the asymptotic behavior of the transverse
difference correlation function is given by
\begin{equation}
  C({\bf x})=\frac{1}{N}\overline{\langle [\bphi(\bx)-\bphi(\bN)]^2
    \rangle}\sim\ln(|{\bf x}|/L_R)
\end{equation}
on both sides of the RT, see Fig.~\ref{fig.mani_flow}(d).  Beyond
$\xi_\|$ the roughness crosses over to the power law 
$C({\bf x})\sim (|{\bf x}|/\xi_\|)^{2\zeta}$ in the rough phase.  
On the flat side of
the transition $C({\bf x})$ saturates on scales larger than $\xi_\|$
at a finite value $\sim \ln(\xi_\|/L_R)$.

The phase boundary between the rough and the flat phase can be
obtained from the condition that the primary solution of Eq.
(\ref{rg2_v}), neglecting the $\bar v^3$ term, and the fixed point
value $\bar v^*$ match at the cross-over length $l_R$. This condition
leads to the result
\begin{equation}
  \label{boundary}
  v_{0,c}(\Delta_0,\hat\xi_0)=g(\epsilon)a_0^2\left(\frac{\hat\xi_0}{a_0}
\right)^{2/\zeta} \left(\frac{\Delta_0}{\hat\xi_0^2}\right)^{2/\epsilon},
\end{equation}
which is in agreement with the estimate of Eq. (\ref{v0c}) derived
from scaling arguments for random field disorder. The coefficient
$g(\epsilon)$ is given to lowest order in $\epsilon$ by
\begin{equation}
  \label{g_coeff}
g(\epsilon)=\frac{\epsilon}{2}\left(\frac{\epsilon N}{\alpha+4\beta}
\right)^{1/\zeta-2/\epsilon}(4e)^{-1/\zeta}.
\end{equation}
In both expressions above, $\zeta$ is always the exponent of the rough
phase for $\hat\xi_0\ll a_0$ and $\zeta=\epsilon/2$ for $\hat\xi_0\gg
a_0$. As can be seen easily from the behavior of $g(\epsilon)$, the
threshold $v_{0,c}$ vanishes exponentially fast for $\epsilon \to 0$.
This is in agreement with the fact that in four dimensions an
arbitrarily small lattice potential leads to a flat phase due to the
sub-logarithmic roughness for $v\equiv 0$, cf. Eq. (\ref{sub_ln}).
The schematic RG flow for $2<D<4$ is shown together with that expected
from the scaling arguments of Section 3 for $D \le 2$ in
Fig.~\ref{fig.mani_flow}.

\subsection{Interfaces in Dipolar Systems} 

As shown in Section 3, in the physically interesting case $D=2$ the
manifold is rough beyond the cross-over length $L_R$ even for a finite
crystal potential. Although $L_R$ is exponentially large for weak
disorder, there exists, strictly speaking, no roughening transition
for $D=2$. However, the transition is expected to be seen even for
$D=2$ in systems with dipolar interactions as already noted in Ref.
\cite{Bouchaud+92b}. In these systems ferroelectric or magnetic domain
walls ($N=1$) separate regions of opposite polarization or
magnetization. Assuming that the dipoles interact via a
three-dimensional Coulomb force and that the dipolar axis $\rho$ is
parallel to the wall, one obtains for $D>1$ in harmonic approximation
the modified elastic energy \cite{Lajzerowicz80,Nattermann83}
\begin{equation}
  \label{dipolar}
  E_{\rm el}=\frac{\gamma}{2}\int\frac{d^D{\bf q}}{(2\pi)^D}
  q^2\left(1+g\frac{q_\rho^2}{q^{D+1}}\right)
  \phi_{-{\bf q}}\phi_{\bf q},
\end{equation}
where $g$ measures the relative strength of the dipolar interaction.
As can be easily seen from Eq. (\ref{dipolar}), the dipolar
interaction increases the elastic stiffness of the interface on large
scales. As a result, the upper critical dimension $D_c$ should
decrease compared to the case of an isotropic elastic kernel $\sim
q^2$ with $D_c=4$.

Indeed, our results can also be applied to interfaces with the elastic
interaction of Eq. (\ref{dipolar}). For the RG analysis of dipolar
interfaces, there are mainly two differences compared to the usual
case discussed before. First, the large scale behavior of the
interface is described correctly if we take the stiffness constant to
be an effective $\bq$-dependent function
$\tilde\gamma(\bq)\sim\sqrt{g}\gamma |\bq|^{(1-D)/2}$.  This can be
seen by writing $q_\rho=q\cos(\varphi)$ and performing the angular
integration over $\varphi$ in the Fourier representation of the free
two-point function. This integration yields an additional factor $\sim
|\bq|^{(D-1)/2}$ in the limit $|\bq| \to 0$ which can be absorbed in
$\tilde\gamma$. Repeating the RG analysis with this modified
$\tilde\gamma$, the form of the terms in the flow equations arising
from scale changes remains unchanged if we replace $\epsilon=4-D$ by
\begin{equation} 
  \label{eps_dipolar}
  \epsilon=\frac{3}{2}(3-D).
\end{equation}
The second difference arises from the momentum shell integral over the
anisotropic elastic kernel. This leads only to a modified numerical
coefficient $K_D$ in the non-linear parts of the flow equations, but
has no influence on the exponents. Therefore, the exponents calculated
above remain valid to first order in $\epsilon$ if we use Eq.
(\ref{eps_dipolar}). The upper critical dimension is shifted to
$D_c=3$, and there exists a roughening transition for two dimensional
dipolar interfaces which is described by our results with
$\epsilon=3/2$.

\subsection{Periodic Elastic Media}

Having established the existence of a roughening transition for single
elastic manifolds, we now turn to the case of periodic elastic media.
From the analysis of Section 5.2 we know that the correlation function
$C(\bx)$ behaves logarithmically for $|\bx|>L_\Delta$ in the rough
unlocked phase (U) described by the stable $T=0$ fixed point $U^*$.
Therefore, one can expect a logarithmic scaling of $\bphi$ also at the
roughening transition, hence we set in this section $\zeta=0$ from the
very beginning.  To extend the RG analysis to $V_L(\bphi) \neq 0$, we
follow the treatment presented above for elastic manifolds. In the
following, we will focus on the particular case of a square lattice
for the elastic medium. Then we are able to make use of the reduction
of the $R(\bphi)$-flow to the single parameter flow for $\Delta$
derived in Section 5.2, cf. Eq. (\ref{Delta_flow}). Again, a finite
lattice potential changes this equation only by a shift in $\epsilon$.
For simplicity, we set the periodicity of the elastic medium to
$l=2\pi$ in the following such that $p=2\pi/a_0$ on all length scales
since transversal lengths are not rescaled here. Then the scaling
equations for elastic media read
\begin{eqnarray}
\label{rg_pm_T}
\frac{dT}{dl}&=&(2-D-\frac{1}{2}p^6\Delta v^2)T,\\
\label{rg_pm_D}
\frac{d\Delta}{dl}&=&\epsilon \Delta - \frac{9}{\pi^2}\Delta^2 
-p^6v^2\Delta^2,\\
\label{rg_pm_v}
\frac{dv}{dl}&=&\left(2-\frac{p^2}{2}\Delta\right)v-\frac{7}{8}
p^6\Delta v^3.
\end{eqnarray}
From the scaling arguments presented in Section 3, one may expect that
for $p<p_c=6/\pi\sqrt{\epsilon}$ weak disorder is always irrelevant
due to the logarithmic roughness in the phase U. For $p>p_c$ a weak
periodic potential is irrelevant, but one expects that it becomes a
relevant perturbation once $v_0$ exceeds a critical value $v_{0,c}$.
Therefore, $v_{0,c}$ has to vanish if $p \rightarrow p_c^+$, and a new
unstable fixed point corresponding to a roughening transition should
be accessible perturbatively in $\epsilon=4-d$ and the second small
parameter $\mu=p^2/p_c^2-1$.

\begin{figure}[htb]
\begin{center}
\leavevmode
\epsfxsize=0.7\linewidth
\epsfbox{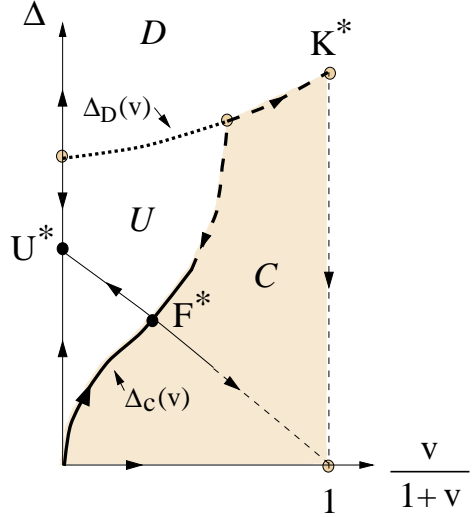}
\end{center}
\caption{Schematic RG flow for $p>p_c$ in
  the $v$-$\Delta$-plane. A critical line $\Delta_c(v)$ with fixed
  point $F^*$ separates the flat ({\it C}) and rough, unlocked ({\it
    U}) phase.  For $p<p_c$ the fixed points $F^*$ and $U^*$ merge.
  Since for large disorder strength $\Delta$ topological defects will
  proliferate, we expect a disordered phase {\it D} above the line
  $\Delta_D(v)$ as will be explained in Section 8.  While the RG
  flow in this range is not yet clear, the transition from the {\it C}
  to the {\it D} phase is probably in the universality class of the
  random field $p$-state clock model.}
\label{fig.media_flow}
\end{figure}

Indeed, there exists a new unstable $T=0$ fixed point $F^*$ which
approaches the stable fixed point describing the rough phase if
$\mu\to 0$. It is parameterized to lowest order in $\epsilon$ and
$\mu$ by
\begin{equation}
  \label{pm_fp_values}
  \Delta^*=\frac{\pi^2}{9}\epsilon(1-\mu),\quad p^2v^*=
\frac{\sqrt{\epsilon\mu}}{2}.
\end{equation}
Linearization around this fixed point gives the eigenvalues
$\lambda_1=-\epsilon$ and $\lambda_2=4\mu$. Thus, one has to identify
this unstable fixed point with the roughening transition. The
longitudinal correlation length $\xi_\|$ beyond which the periodic
medium is locked into the minima of the lattice potential diverges
according to
\begin{equation}
  \label{pm_xi}
  \xi_\|\sim |v-v_{0,c}|^{-\nu_\|}, \quad \nu_\|=\frac{1}{4\mu}
\end{equation}
if the transition is approached from the flat side. On the rough side 
of the transition $\xi_\|$ is infinite, in contrast to the situation
for single elastic manifolds.  The resulting phase diagram is
characterized by a critical line $\Delta_c(v)$ separating the flat or
commensurate (C) and rough, unlocked phase (U), see Fig.
\ref{fig.media_flow}. The position of this critical line is given by
the scaling result $v_{0,c}\approx p^{-2}\Delta_0^{2/\epsilon}$
derived in Section 3. 

To characterize the order of the periodic medium, we introduce the
translational order parameter
\begin{equation}
\Psi(\bx)=\rho_1 \exp(i\phi(\bx)).
\end{equation}
For simplicity, we study here the case of a scalar field ($N=1$) 
only. Inside the C phase, one has a nonzero expectation value
$\overline{\langle \Psi(\bx)\rangle}$. The correlation function of the
fluctuations $\delta\Psi(\bx)=\Psi(\bx)-\overline{\langle 
  \Psi(\bx)\rangle}$ decays exponentially in this phase, 
\begin{equation}
  \label{psi_corr}
  K(\bx)=\overline{\langle \delta\Psi(\bx) \delta\Psi^*(\bN) \rangle}
  \sim \exp(-|\bx|/\xi_\|).
\end{equation}
Approaching the transition from the C phase, the order parameter
vanishes continuously as
\begin{equation}
  \label{exp_beta}
  \overline{\langle \Psi(\bx)\rangle}\sim (v_0-v_{0,c})^\beta,\quad
  \beta=\frac{\pi^2}{18}\epsilon\, \nu_\|.
\end{equation}
In contrast, at the transition and in the rough phase, $v_0\le
v_{0,c}$, the periodic medium is unlocked with $\overline{\langle
  \Psi(\bx)\rangle}=0$ and the correlation function shows a power law
decay,
\begin{equation}
  \label{psi_corr_rough}
  K(\bx)\sim |\bx|^{4-D-\bar\eta}, \quad \bar\eta=(1+\pi^2/9)\epsilon.
\end{equation}
At the transition the 'violation of hyperscaling' exponent is
increased from the exact value $\theta=D-2$ in the rough phase to
\begin{equation}
  \label{pm_theta_c}
  \theta_c=D-2+\epsilon\mu/2
\end{equation}
due to the renormalization of the elastic stiffness of the medium.
The critical exponents mentioned above fulfill all scaling relations
and exponent inequalities which hold for random field models with a
discrete symmetry of the order parameter
\cite{Villain85,Nattermann+88a}.
 
\subsection{The Limit $N=\infty$}

It is important to note that the order as well as the critical
exponents $\nu_\|$ and $\zeta_c$ of the roughening transition do not
depend on $N$ as long as $N$ is finite.  For arbitrary $N$, the
roughening transition of manifolds in random bond systems has also
been studied within a variational approach by Bouchaud and Georges
(BG) \cite{Bouchaud+92b}.  At $T=0$ and for $\hat\xi_0\ll a_0$ these
authors find a first order transition between a glassy rough and a
glassy flat phase.  In the glassy rough phase they obtain the Flory
result $\zeta=\epsilon/5$ which is known to be not correct for small
$\epsilon$. We attribute this as well as the first order nature of the
transition to the use of the variational method which is known to give
spurious first order transitions \cite{Patashinsky+79}. However, the
variational approach is expected to become exact for $N=\infty$.
Since our results seem to be in contrast to the $N=\infty$ results of
BG, we will study the RG flow for elastic manifolds in this limit now.

Our discussion starts again with the flow Eqs.
(\ref{rg2_D})-(\ref{rg2_v}). To have a finite bare starting value for
$\bar \xi$ in the limit $N\to \infty$, we rescale this parameter
according to $\bar \xi^2 \to N \bar \xi^2$. For random field disorder,
the structure of the flow equations remains unchanged in the limit
$N\to \infty$ since $\beta=N/2$. Therefore, the transition is still of
second order and the phase boundary is given by Eq.  (\ref{boundary})
with the $N$-independent coefficient of Eq.  (\ref{g_coeff}). In
contrast, for random bond disorder the manifold is only
logarithmically rough in the case $N=\infty$, cf. Eq.
(\ref{rb_ln_corr}), and we expect a roughening transition similar to
that of periodic media. Indeed, due to the above rescaling of $\bar
\xi$ and $\beta=1/2$ we obtain $d\bar\xi^2/dl=0$ and, therefore, Eq.
(\ref{rg2_D}) reduces to
\begin{equation}
\label{N_infty_flow}
\frac{d\bar\Delta}{dl}=\epsilon\bar\Delta-\left[\frac{1}{2p^2}+\bar
  v^2 \right]\bar\Delta^2,
\end{equation}
where $p=2\pi \hat\xi_0/a_0$ is a constant. Comparing this equation
with the flow Eq. (\ref{rg_pm_D}) of $\Delta$ for periodic media, we
conclude that for $p<p_c=\sqrt{2/\epsilon}$ the manifold is already in
the flat phase for an infinitesimally small $v_0$. A roughening
transition at a finite $v_0$ exists only for $p>p_c$, and then the
correlation length diverges as described by Eq. (\ref{pm_xi}).

The variational calculation of BG applies in the random bond case
where $\hat\xi_0 \ll a_0$, i.e. for $p<p_c$. Therefore, for $N=\infty$
we expect no transition at a finite $v_0$. Indeed, this is in
agreement with the result for the phase boundary of BG in the limit
$N=\infty$.  The resulting transition at $v_{0,c}=0$ for $p<p_c$ is
not described by a fixed point of our RG equations and we have been
unable to ascertain definitely the nature of the transition in this
case. We will discuss this in more detail in Section 8 in the context
of periodic media.

\subsection{Incommensurate Phases}

So far we have considered only periodic elastic media with a
periodicity $l=pa_0$ which is exactly an integer multiple of the
crystal lattice periodicity $a_0$. But in general, this condition is
not fulfilled exactly in real systems and one has to assume a shifted
lattice constant
\begin{equation}
\label{def_misfit}
l=pa_0(1-pa_0\delta)
\end{equation}
for the elastic medium, where $\delta$ is a small misfit. For example,
$\delta$ can be changed in flux line lattices by tuning the external
magnetic field or in charge density waves by changing the temperature.
It is well known from the sine-Gordon model subject to thermal
fluctuations that above some critical value $\delta_c$ of the misfit,
the incommensurability is compensated by the formation of domain walls
or solitons of intrinsic width given by the correlation length
$\xi_\|$ \cite{Pokrovsky+79}. Across these walls the phase field
$\phi(\bx)$ changes by $2\pi/p$. In the following, we specialize our
discussion of the incommensurate phases to the case of a scalar field
($N=1$).

The effect of a finite misfit can be described by adding a linear
gradient term to the Hamiltonian defined in Eq. (\ref{genHam}), such
that the elastic term is replaced by
\begin{equation}
  \label{lin_grad}
  \frac{\gamma}{2} \int d^D \bx \left\{ \bbox{\nabla} \phi
    -\bbox{\delta}\right\}^2
\end{equation}
with $\bbox{\delta}=\delta \hat{\bf z}$ and $\hat{\bf z}$ the
modulation direction of the periodic medium. For completeness we note
that for a vanishing lattice potential $V_L$ the misfit $\delta$ can
be eliminated by the transformation $\phi(\bx)\to \phi(\bx)+\delta z$
due to the statistical symmetry of the disorder under a shift in
$\phi$. The linear gradient term in Eq. (\ref{lin_grad}) can be
integrated out leading to the contribution
\begin{equation}
  \label{sol_density}
  -\gamma\delta\int d^D \bx \frac{2\pi}{p} \frac{1}{\ell}
\end{equation}
to the Hamiltonian (\ref{genHam}), where we have introduced the mean
soliton density $\ell^{-1}=p[\phi(z=L)-\phi(z=0)]/2\pi L$ for a system
of size $L$ in $\hat{\bf z}$-direction.

\begin{figure}[htb]
\begin{center}
\leavevmode
\epsfxsize=0.9\linewidth
\epsfbox{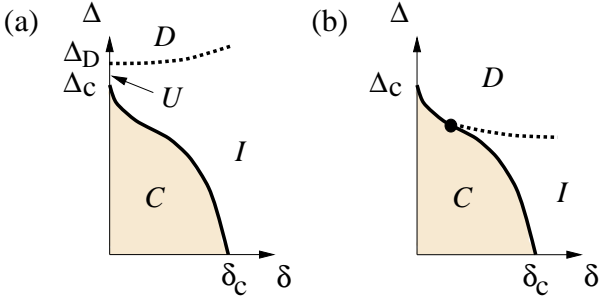}
\end{center}
\caption{Schematic phase diagrams for finite misfit $\delta$ at fixed
  crystal potential strength $v_0$ for $p>p_c$ (a) and $p<p_c$
  (b). Due to the proliferation of topological defects a disordered
  phase is expected above a disorder strength $\Delta_D$.}
\label{fig_ic}
\end{figure}

To study the case of finite lattice pinning, a direct RG treatment of
the Hamiltonian (\ref{genHam}) with an additional term
(\ref{sol_density}) is not justified since a finite soliton density
$\ell^{-1}$ induces linearly growing contribution $\sim \ell^{-1} z$
to the field $\phi(\bx)$ which cannot be Fourier expanded. To avoid
this problem, we go back to periodic boundary conditions by the
transformation $\phi(\bx)\rightarrow \phi(\bx)-2\pi z/p\ell$
\cite{Horovitz+83}. Note that we treat the mean soliton distance
$\ell$ as a fixed parameter which is not renormalized and has to be
determined later by minimizing the free energy of the soliton lattice.
In the commensurate phase (C) for $\delta<\delta_c$, the soliton
density vanishes and hence the RG analysis for $\delta=0$ applies. To
estimate $\delta_c$, we compare the energy gain of a soliton per unit
area, $-\gamma\delta 2\pi/p$, with its cost for vanishing misfit,
$\gamma\sqrt{v_{\rm eff}}2\pi/p$, where $v_{\rm eff}=(p\xi_\|)^{-2}$
denotes the effective strength of the lattice potential renormalized
by disorder.  This leads to the result
\begin{equation}
  \label{delta_c}
  \delta_c\approx \frac{1}{p\xi_\|}=\sqrt{v_0}
  \left(1-\frac{\Delta_0}{\Delta_{0,c}}\right)^{\nu_\|},
\end{equation}
where we have used Eq. (\ref{pm_xi}). The corresponding phase diagram
for fixed $v_0$ is shown in Fig. \ref{fig_ic}. Since for $p<p_c$ the
lattice potential is always relevant for weak disorder, there exists
in this case for large enough disorder a direct transition to a
disordered phase (D). This transition is not described by our approach
since it is driven by the proliferation of topological defects as will
be explained in Section 8. In deriving expression (\ref{delta_c}), we
have neglected a possible additional reduction of $\delta_c$ by
fluctuations of the solitons on scales beyond $\xi_\|$ which have not
been integrated out by the RG.

In the incommensurate phase (I) the above transformation to periodic
boundary conditions induces a spatial oscillation in the amplitude of
the lattice potential,
\begin{equation}
  \label{V_L_osci}
  V_L(\phi)=v_0 \cos(p\phi-2\pi z/\ell).
\end{equation}
Notice that the same oscillating pinning potential describes vicinal
surfaces which are tilted by a small angle $\theta=2\pi/p\ell$ with
respect to the crystal planes \cite{Nozieres92}. In the RG calculations,
this leads to an additional factor $\exp(i2\pi z/\ell)$ in the
integral of Eq. (\ref{kernel_eval}). When the lattice potential is
relevant, $\Delta_0 < \Delta_{0,c}$, there are two length
scales: $\ell$ and the correlation length $\xi_\|$. 

If $\xi_\|\ll \ell$, the RG up to $\xi_\|$ is insensitive to the
oscillations along the $z$-direction and the soliton lattice can be
described in terms of an effective free energy.  The corresponding
free energy density is given by \cite{Nattermann83}
\begin{equation}
  \label{soliton_energy}
  f(\ell)\simeq\frac{1}{\ell}\left\{\sigma\left(1-\frac{\delta}{\delta_c}
      \right)+4\sigma e^{-\ell/\xi_\|}+B(D)\sigma\ell^{-1/\bar\beta}\right\}.
\end{equation}
Here the first term is the free energy of a soliton per unit area, the
second term is the bare soliton repulsion, and the last part describes
the steric repulsion caused by collisions of the solitons. The
effective soliton tension is $\sigma=8\gamma/p^2\xi_\|$, $B(D)$ is a
coefficient which vanishes for $D>5$ and
$\bar\beta=\zeta/(2-2\zeta)=(5-D)/2(D-2)$ where $\zeta=(5-D)/3$ is the
roughness exponent of the soliton walls subject to random field
disorder.  Minimizing the free energy, we obtain the mean soliton
distance near the IC-transition
\begin{equation}
  \label{mean_ell}
  \ell\sim (\delta-\delta_c)^{-\bar\beta}.
\end{equation}

To determine the translational order inside the I phase, the soliton
lattice itself can be considered, on scales larger than $\ell$ as a
periodic elastic medium with a disorder correlator $R_0(\phi)$ of
periodicity $\ell$. The corresponding Hamiltonian is given by Eq.
(\ref{genHam}) with $V_L=0$ and $N=1$ but with anisotropic elasticity.
The elastic constants are $\gamma_\|=\sigma/\ell$ parallel to the
solitons and $\gamma_\perp=\ell^2 f''(\ell)$ in $\hat{\bf z}$
direction. Rescaling the $z$ coordinate according to
$\sqrt{\gamma_\|/\gamma_\perp}z \to z$, we can achieve again an
isotropic elastic term with $\gamma=\sqrt{\gamma_\|\gamma_\perp}$.
Using the result of Eq. (\ref{corr_pm}) we obtain a logarithmically
diverging correlation function for the soliton displacement field
$u(\bx)$,
\begin{equation}
  \label{u_corr}
  \overline{\langle [u(\bx)-u(\bN)]^2\rangle}=\frac{4-D}{36}\ell^2
  \ln\left(\frac{\bx_\|^2}{L_\|^2}+\frac{z^2}{L_\perp^2}\right).
\end{equation}
Here the Larkin lengths $L_\|\sim (\ell^4 \gamma_\|^{3/2}
\gamma_\perp^{1/2})^{1/(4-D)}$ and
$L_\perp=\sqrt{\gamma_\perp/\gamma_\|}L_\|$ are the correlations
length beyond which the logarithmic divergence becomes asymptotically
exact. Since $\gamma_\|\sim \ell^{-1}$ and $\gamma_\perp \sim
\ell^{(\zeta-2)/\zeta}=\ell^{-(D+1)/(5-D)}$ for random fields, both
correlation lengths diverge if the IC-transition is approached from
the I phase as
\begin{eqnarray}
  \label{larkin_div}
  L_\| &\sim& \ell^{3/(5-D)} \sim (\delta-\delta_c)^{-3/2(D-2)}\\
  L_\perp &\sim& \ell \sim (\delta-\delta_c)^{-(5-D)/2(D-2)}
\end{eqnarray}
similar to the situation for the IC-transition in the presence of
thermal fluctuations only \cite{Nattermann83,Huse+84}. 

The behavior of the correlation function $K(\bx)$ of the fluctuations
of the translational order parameter $\Psi(\bx)$ can be easily
obtained from Eq. (\ref{u_corr}). The soliton lattice shows quasi long
range order with
\begin{equation}
\label{algedec}
K(\bx)=\rho_1^2\cos{\frac{2\pi z}{p\ell}\left(
\frac{\bx_\|^2}{L_\|^2} + \frac{z^2}{L_\perp^2}
\right)^{-(4-D)\pi^2/18p^2}}.
\end{equation}
Here we have used that in the I phase a difference in the soliton
displacement is related to a shift of the phase $\phi(\bx)$ of the
order parameter by
$\phi(\bx)-\phi(\bN)=(2\pi/p)(z-u(\bx)+u(\bN))/\ell$.

In the opposite case where $\xi_\| \gg \ell$, the lattice potential
$V_L$ should be irrelevant since it averages to zero on scales larger
than $\ell$ for geometric reasons. This corresponds to the situation
above the roughening transition, $\Delta_0>\Delta_{0,c}$, and now the
original periodic medium (instead of the soliton lattice) is described
by the Hamiltonian (\ref{genHam}) with $V_L=0$ since the misfit can
now be shifted away as mentioned above. Therefore, the correlation
function has to show a crossover to a behavior which is given by Eq.
(\ref{algedec}) where the Larkin lengths of the soliton lattice become
that of the original periodic medium, $L_\|$, $L_\perp \to L_\Delta$.
The exponent of the algebraic decay in Eq. (\ref{algedec}) goes over
to $-(4-D)\pi^2/18$.

\section{Experimental Implications}

Having established the possibility of a disorder driven roughening
transition for various systems which can be described in terms of
elastic objects, we discuss now the experimental consequences of our
results. To begin with, we comment on how the transition may be
approached for a sample with {\it fixed} disorder strength. In
general, for both elastic manifolds and periodic elastic media one may
expect that the bare pinning strengths $v_0$ and $\Delta_0$ depend
also on temperature.  Hence the transition should be reachable by
changing the temperature $T$ for appropriate $T$-dependencies of both
pinning strengths. The precise $T$-dependence of the parameters is
model dependent and, therefore, we will give only some general
arguments and typical examples. Close to the condensation temperature
$T_c$ of the system where the intrinsic width of the elastic manifold
becomes much larger than the periodicity of the lattice, $\xi_0 \gg
a_0$, we expect a very weak influence of the crystal potential. An
example are domain walls in magnets where $T_c$ is the Curie
temperature. For Peierls barriers which appear, e.g., for domain walls
in ferroelastics or solitons in incommensurate phases, one expects an
exponentially decreasing influence of the lattice potential, $v_0\sim
e^{-C\xi_0}$.

The most prominent example for periodic elastic media are charge
density waves. In this case the critical ratio of the pinning
strengths depend also on temperature since $v_0/\Delta_0^2 \sim
\rho_1^{2+p}$ for $D=3$ where the charge density amplitude $\rho_1$
changes with $T$ and vanishes at the Peierls temperature.  Thus, for
not too large disorder, we can expect to see the roughening transition
by increasing $T$ for both elastic manifolds and periodic media.

Indications about which phase is present should be given by the creep
behavior since at small driving forces the dynamics are governed by
the static behavior of the elastic objects. Indeed, under the
influence of a small driving force density $f_{\rm ex}$, the motion of
the elastic object is dominated by jumps between neighboring
metastable states in the rough phase and between adjacent minima of
the lattice potential in the flat phase. As has been discussed earlier
\cite{Ioffe+87,Nattermann90,Blatter+94}, in both cases the creep
velocity $u$ is exponentially small,
\begin{equation}
\label{creep}
u(f_{\rm ex})\sim \exp\left[-\frac{E_c}{T}\left(\frac{f_c}
{f_{\rm ex}}\right)^\kappa\right],
\end{equation}
where $f_c$ denotes the maximal pinning force density of the
dominating pinning mechanism and $E_c$ the corresponding energy
barrier. The creep exponent $\kappa$ depends on the phase which is
present and on the type of disorder under consideration, i.e. on the
disorder correlator $R_0(\bphi)$.

For periodic elastic media, we obtain in the rough unlocked and flat
phase, respectively,
\begin{equation}
\label{kappa_pm}
\kappa_{\rm rough}=\frac{D-2}{2}, \quad \kappa_{\rm flat}=D-1.
\end{equation}
The pinning force density and the energy barrier in the rough phase
are given by $f_c\approx \gamma L_\Delta^{-2}$ and $E_c\approx \gamma
L_\Delta^{D-2}$, whereas in the commensurate flat phase they are 
$f_c\approx\gamma\xi_\|^{-2}$ and $E_c\approx \gamma\xi_\|^{D-2}$.

For elastic manifolds, the exponent $\kappa$ depends on the type of
disorder via the roughness exponent $\zeta$ , i.e. random bond or
random field disorder,
\begin{equation}
  \label{kappa_em}
  \kappa_{\rm rough}=\frac{D-2+2\zeta}{2-\zeta}, \quad
  \kappa_{\rm flat}=D-1.
\end{equation}
In the rough phase, we obtain $E_c\sim \gamma\xi_0^2 L_R^{D-2}$ and
$f_c\sim \gamma\xi_0 L_R^{-2}$, and in the flat phase $E_c\sim \gamma
v^{(2-D)/2}$ and $f_c\sim \gamma v$ for an interface with short range
interactions.  For dipolar systems one has to replace the dimension
$D$ according to Eq. (\ref{eps_dipolar}), which gives $\kappa^{RF}=1$,
$\kappa^{RB}=0.6666$ for a two dimensional interface in the rough
phase and $\kappa=1$ for a flat interface independent of the type of
disorder.  Therefore, in the RB case the phase of the interface can be
determined by measuring $\kappa$.

A very recent experiment on driven domain walls \cite{Lemerle+98}
shows that the exponent $\kappa$ can be measured very accurately. This
should make it possible to distinguish experimentally between both
phases in the random bond case. For periodic media, an experimental
realization is given by charge density waves. Recently measured I-V
curves of the conductor o-TaS$_3$ at temperatures below 1K can be
fitted by Eq. (\ref{creep}) with $\kappa=1.5$ -- $2$ \cite{Zaitsev+97}.
The experimentally observed tendency to larger $\kappa$ for purer
crystals is in agreement with Eq. (\ref{kappa_pm}) for $D=3$.

\section{Discussion and Conclusions}

In this paper we have studied the static behavior of oriented elastic
objects subjected to both a random potential and a periodic potential
from an underlying crystal lattice. Scaling arguments and a functional
RG calculation in $D=4-\epsilon$ dimensions led us to the conclusion
that the competition between both pinning effects induces a {\it
  continuous} roughening transition between a flat phase with finite
displacements and a rough state with diverging displacements. While
based on the same physical mechanism, the description of the
transition turned out to be different for periodic elastic media and
single elastic manifolds.

For periodic elastic media such as charge density waves or flux line
lattices in high-$T_c$ superconductors, the transition exists only
above some critical ratio $p_c=l/a_0$ of the lattice constants of the
medium and the crystal potential. In $D=3$ we obtain $p_c=6/\pi$ from
the $\epsilon$-expansion. Below $p_c$, a weak random potential is
irrelevant and the medium remains flat.  On the flat side, this
transition is described by three independent critical exponents which
obey the scaling relations for random field systems with a discrete
symmetry of the order parameter. Although the situation is similar to
the thermal roughening transition in two dimensions, where also
logarithmic roughness exists in the rough phase, in the disorder
driven case the existence of one universal stable fixed point
describing the rough phase leads to a different RG flow. In contrast,
for the thermal case the temperature axis represents a line of fixed
points. Notice that in the disorder dominated system studied here, the
important logarithmic roughness sets in only beyond the Fukuyama-Lee
length $L_\Delta$, whereas thermal roughness in two dimensions appears
on all length scales.

In the case of elastic manifolds, the interplay between random
potential and lattice pinning leads also to a {\it continuous}
transition for $2<D<4$. We have obtained the exponent
$\nu_\|=1/2\sqrt{\epsilon}$ for the divergence of the correlation
length, which has to lowest order in $\epsilon$ the same form as that
of the thermal roughening transition if $\epsilon=2-D$.  For both
random field and random bond systems, elastic manifolds show a
superuniversal logarithmic roughness {\it at} the transition.

For $D=2$ the interface is always asymptotically rough. At the upper
critical dimension $D=4$ an arbitrarily small lattice pinning produces
a flat interface due to sub-logarithmic roughness in the absence of
lattice effects. A diverging correlation length $\xi_\|$ appears on
both sides of the transition. On the rough side $\xi_\|$ sets the
length scale for the crossover from logarithmic to algebraic
roughness.

Recently, the exact ground state of interfaces in the three
dimensional random bond, cubic lattice Ising model has been studied
numerically \cite{Alava+96}. For interfaces oriented along the
$\{100\}$ direction, a roughening transition at finite disorder
strength from an almost flat to a rough phase has been observed
whereas interfaces along the $\{111\}$ direction remain always rough.
Although our scaling arguments yield even for interfaces along the
$\{100\}$ direction no transition, one has to take into account the
finite system sizes used in the numerical calculation. Indeed, we
expect a flat behavior of the interface width up to the scale $L_R$,
which is exponentially large for weak disorder in $D=2$, hence
probably leading to a flat 'phase' at sufficiently weak disorder in
finite systems.

More recently, also the replica Hamiltonian for periodic media with
$R_0(\phi)\sim \cos(\phi)$, $N=1$, has been studied using the Gaussian
variational method for $p=1$, $2\le D<4$ \cite{Sasada96} and for
$p=1/2$, $D=3$ \cite{Fridrikh+97}. Both analysis yield a
self-consistency equation for the effective mass of the field
$\phi(\bx)$ in terms of the disorder strength.  Solutions of this
equation exist only for $v_0$ above a particular value $v_{0,c}$,
which is then identified with the phase boundary. At this phase
boundary the effective mass remains {\it finite} corresponding to a
first order transition.  Moreover, the threshold $v_{0,c}$ is
non-vanishing in spite of $p<p_c$. As we have shown by scaling
arguments and the RG calculation, these results cannot be correct in
the case of weak disorder.  To understand the origin of these
inconsistent results, we note that the derivation of the implicit
equation is based on the assumption of a replica symmetric solution
for the rough phase, hence corresponding to the perturbative result of
Larkin. In the rough phase, for the correct logarithmic roughness on
scales beyond the Larkin length $L_\Delta$, replica symmetry breaking
is needed within the variational approach. Since this logarithmic
scaling on asymptotic scales is essential for the existence of a
critical $p_c$, the replica symmetric determination of the effective
mass cannot reproduce our RG result. In addition, the self-consistency
equation cannot expected to be valid in the vicinity of a second order
transition since it does not capture the physics on scales $\xi_\| \gg
L_\Delta$ correctly.  On the other hand, it should be noted that the
variational approach, in contrast to the RG, does not depend on the
assumption of weak pinning, hence describing the system perhaps
correctly in the case of strong pinning. However, the variational
approach remains an uncontrolled approximation.

For $p<p_c$, the correlation length $\xi_\|$ has to diverge for $v_0
\to v_{0,c}=0$.  Now we address the question of the corresponding
exponent $\nu_\|$.  For this transition, there exists no unstable
fixed point of the RG equations. As a first step towards an answer of
this problem, one can calculate self-consistently an effective mass.
To extend the validity range of the self-consistency equation derived
in Refs.  \cite{Sasada96,Fridrikh+97} to weak disorder, we have taken
into account the correct asymptotic roughness by using the RG result
for the case where $v_0=0$ . Then the solution for the effective mass
depends on $p$. For $p<p_c$ the mass remains always finite for $v_0>0$
and vanishes {\it continuously} as $v_0=0$ is approached, leading to
the exponent $\nu_\|=1/2(1-p^2/p_c^2)$ \cite{Emig+98b}. If $p>p_c$,
the mass remains finite until $v_{0,c}>0$, where a solutions ceases to
exist. In the latter case our RG treatment applies. We believe that
the second order nature of the transition for $p<p_c$ is correctly
given by our self-consistency condition, but the exponent $\nu_\|$
estimated from this condition should not be taken seriously.

\begin{figure}[htb]
\begin{center}
\leavevmode
\epsfxsize=1.0\linewidth
\epsfbox{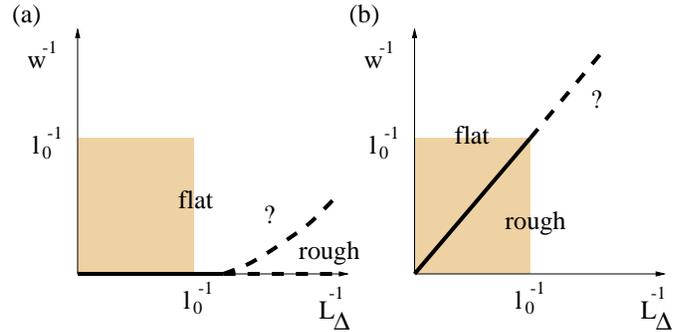}
\end{center}
\caption{Speculative phase diagrams including also the case of strong pinning
  where $w\sim v_0^{-1/2}$ and $L_\Delta\sim \Delta_0^{-1/(4-D)}$
  become of the order of $l_0=$max($a_0$,$\xi_0$) for (a) $p<p_c$ and
  (b) $p>p_c$. In the shaded region the RG is expected to be valid.
  Dashed lines represent possible locations of the roughening
  transition which may be of first order outside the validity region of
  the RG.}
\label{fig_specu}
\end{figure}

The preceding discussion can be summarized in a speculative phase
diagram, which includes also the strong pinning case, see Fig.
\ref{fig_specu}.  If the pinning potentials are strong enough such
that the characteristic scales $w\sim v_0^{-1/2}$ and $L_\Delta \sim
\Delta_0^{-1/(4-D)}$ become of the order of max($a_0$,$\xi_0$), a
first order transition may appear.

In our study of periodic media we have not attempted to include the
effect of topological defects. These can be considered if we treat
$\phi({\bf x})$ as a multi-valued field which may jump by $\pm 2\pi$ at
surfaces which are bounded by vortex lines.  Including these defects,
qualitatively we expect the following picture. For $\delta=0=v_0$ it has
been argued recently that for weak enough disorder strength
$\Delta_0<\Delta_D$, the system is stable with respect to the formation
of vortices \cite{Gingras+96,Kierfeld+97,FisherDS97}.  However, vortex
lines will proliferate for $\Delta_0>\Delta_D$. At present it is not
clear whether the corresponding transition is continuous or first
order. For $\delta=0$ but $v_0>0$ we expect that this transition extends
to a line $\Delta_D(v_0)$ until $v_0$ reaches a critical value $v_D$ with
$\Delta_D(v_D)=\Delta_c(v_D)$ (see Fig.  \ref{fig.media_flow}). For
larger $v_0$ the transition is probably in the universality class of the
$p$-state clock model in a random field, which has an upper critical
dimension $d_c=6$. For $p=2$ (Ising like case) the transition is of
second order \cite{Nattermann+89}. A non-zero value of $\delta$ will
in general increase the size of the incommensurate phase, as
schematically sketched in Fig.  \ref{fig_ic}.

\begin{acknowledgement}    
  We acknowledge discussions with L. Balents, S. Scheidl and A.
  Georges.  T.N.\ acknowledges support of the Volkswagen-Stiftung and
  the German-Israeli Foundation (GIF), and T.E.\ acknowledges support
  of the German-Israeli Foundation (GIF) and of the Deutsche
  Forschungsgemeinschaft (SFB 341).
\end{acknowledgement}


\begin{thebibliography}{10}

\bibitem{Chaikin+95}
P.~M. Chaikin and T.~C. Lubensky,
\newblock {\em Principles of condensed matter physics},
\newblock Cambridge Univ. Press, Cambridge, 1995.

\bibitem{Wolf+85}
P.~E. Wolf, F.~Gallet, S.~Balibar, E.~Rolley, and P.~Nozi\`eres,
\newblock J. Physique {\bf 46}, 1987 (1985).

\bibitem{Weeks80}
J.~D. Weeks,
\newblock in {\em Ordering in Strongly Fluctuating Systems}, edited by
  T.~Riste, Plenum, 1980.

\bibitem{vanBeijeren+87}
H.~van Beijeren and I.~Nolden,
\newblock in {\em Structure and Dynamics of Surfaces II}, edited by
  W.~Schrommers and P.~von Blanckenhagen, Springer, Berlin, 1987.

\bibitem{Frohlich54}
H.~Fr\"ohlich,
\newblock Proc. R. Soc. London, Ser. A {\bf 223}, 296 (1954).

\bibitem{Gruner94}
G.~Gr\"uner,
\newblock {\em Density waves in solids},
\newblock Addison-Wesley, Reading, Massachusetts, 1994.

\bibitem{Bak+82}
P.~Bak and R.~Bruinsma,
\newblock Phys. Rev. Lett. {\bf 49}, 249 (1982).

\bibitem{Dvorak79}
V.~Dvorak,
\newblock in {\em Proc. Karpacz Winter School of Theoretical Physics}, edited
  by A.~Pekalski and J.~Przystawa, Berlin, 1979, Springer.

\bibitem{Andrei+88}
E.~Y. Andrei et~al.,
\newblock Phys. Rev. Lett. {\bf 60}, 2765 (1988).

\bibitem{Seshadri+92}
R.~Seshadri and R.~M. Westervelt,
\newblock Phys. Rev. B {\bf 46}, 5142 (1992).

\bibitem{Balents+95b}
L.~Balents and D.~R. Nelson,
\newblock Phys. Rev. B {\bf 52}, 12951 (1995).

\bibitem{Nieber+93}
S.~Nieber and H.~Kronm\"uller,
\newblock Physica C {\bf 210}, 188 (1993).

\bibitem{Fridrikh+97}
S.~V. Fridrikh and E.~M. Terentjev,
\newblock Phys. Rev. Lett. {\bf 79}, 4661 (1997).

\bibitem{Wallace80}
D.~J. Wallace,
\newblock Field theories of surfaces,
\newblock in {\em Gauge Theories and Experiments at High Energies}, edited by
  K.~C. Bowler and D.~G. Sutherland, SUSSP Publications, Edinburgh, 1980.

\bibitem{Montvay+94}
I.~Montvay and G.~M\"unster,
\newblock {\em Quantum Fields on a Lattice},
\newblock Cambridge Univ. Press, Cambridge, 1994.

\bibitem{Chui+76}
S.~T. Chui and J.~D. Weeks,
\newblock Phys. Rev. B {\bf 14}, 4978 (1976).

\bibitem{Knops77}
H.~J.~F. Knops,
\newblock Phys. Rev. Lett. {\bf 39}, 766 (1977).

\bibitem{Jose76}
J.~Jose,
\newblock Phys. Rev. D {\bf 14}, 2826 (1976).

\bibitem{Luther+75}
A.~Luther and I.~Peshel,
\newblock Phys. Rev. B {\bf 12}, 3908 (1975).

\bibitem{Kosterlitz77}
J.~M. Kosterlitz,
\newblock J. Phys. C {\bf 10}, 3753 (1977).

\bibitem{Forgacs+91}
G.~Forgacs, R.~Lipowsky, and T.~M. Nieuwenhuizen,
\newblock The behavior of interfaces in ordered and disordered systems,
\newblock in {\em Phase Transitions and Critical Phenomena}, edited by C.~Domb
  and J.~L. Lebowitz, volume~14, Academic Press, London, 1991.

\bibitem{FisherDS+82}
D.~S. Fisher and M.~E. Fisher,
\newblock Phys. Rev. B {\bf 25}, 3192 (1982).

\bibitem{Halpin+95}
T.~Halpin-Healy and Y.-C. Zhang,
\newblock Phys. Rep. {\bf 254}, 215 (1995).

\bibitem{Villain82}
J.~Villain,
\newblock J. Physique {\bf 43}, L551 (1982).

\bibitem{Grinstein+82}
G.~Grinstein and S.~K. Ma,
\newblock Phys. Rev. Lett. {\bf 49}, 685 (1982).

\bibitem{Binder83}
K.~Binder,
\newblock Z. Phys. B {\bf 50}, 343 (1983).

\bibitem{Villain84}
J.~Villain,
\newblock Phys. Rev. Lett. {\bf 52}, 1543 (1984).

\bibitem{Ioffe+87}
L.~B. Ioffe and V.~M. Vinokur,
\newblock J. Phys. C {\bf 20}, 6149 (1987).

\bibitem{Nattermann87}
T.~Nattermann,
\newblock Europhys. Lett. {\bf 4}, 1241 (1987).

\bibitem{Nattermann+90}
T.~Nattermann, Y.~Shapir, and I.~Vilfan,
\newblock Phys. Rev. B {\bf 42}, 8577 (1990).

\bibitem{Nattermann+88b}
T.~Nattermann and I.~Vilfan,
\newblock Phys. Rev. Lett. {\bf 61}, 223 (1988).

\bibitem{Nattermann+92}
{T. Nattermann et al.},
\newblock J. Physique II {\bf 2}, 1483 (1992).

\bibitem{Nattermann84}
T.~Nattermann,
\newblock Z. Phys. B {\bf 54}, 247 (1984).

\bibitem{Nattermann85}
T.~Nattermann,
\newblock Phys. Stat. Sol. (b) {\bf 132}, 125 (1985).

\bibitem{Bouchaud+92b}
J.~P. Bouchaud and A.~Georges,
\newblock Phys. Rev. Lett. {\bf 68}, 3908 (1992).

\bibitem{Emig+97}
T.~Emig and T.~Nattermann,
\newblock Phys. Rev. Lett. {\bf 79}, 5090 (1997).

\bibitem{Emig+98}
T.~Emig and T.~Nattermann,
\newblock Phys. Rev. Lett. {\bf 81}, 1469 (1998).

\bibitem{Giamarchi+94}
T.~Giamarchi and P.~{Le Doussal},
\newblock Phys. Rev. Lett. {\bf 72}, 1530 (1994).

\bibitem{Giamarchi+95}
T.~Giamarchi and P.~{Le Doussal},
\newblock Phys. Rev. B {\bf 52}, 1242 (1995).

\bibitem{Morgenstern+81}
I.~Morgenstern, K.~Binder, and R.~M. Hornreich,
\newblock Phys. Rev. B {\bf 23}, 287 (1981).

\bibitem{Nattermann90}
T.~Nattermann,
\newblock Phys. Rev. Lett. {\bf 64}, 2454 (1990).

\bibitem{Schulz+88}
U.~Schulz, J.~Villain, E.~Br\'ezin, and H.~Orland,
\newblock J. Stat. Phys. {\bf 51}, 1 (1988).

\bibitem{Balents+93}
L.~Balents and D.~S. Fisher,
\newblock Phys. Rev. B {\bf 48}, 5949 (1993).

\bibitem{Balents+94}
L.~Balents and M.~Kardar,
\newblock Phys. Rev. B {\bf 49}, 13030 (1994).

\bibitem{FisherDS86b}
D.~S. Fisher,
\newblock Phys. Rev. Lett. {\bf 56}, 1964 (1986).

\bibitem{Nattermann+91}
T.~Nattermann and H.~Leschhorn,
\newblock Europhys. Lett. {\bf 14}, 603 (1991).

\bibitem{Lajzerowicz80}
J.~Lajzerowicz,
\newblock Ferroelectrics {\bf 24}, 179 (1980).

\bibitem{Nattermann83}
T.~Nattermann,
\newblock J. Phys. C {\bf 16}, 4125 (1983).

\bibitem{Villain85}
J.~Villain,
\newblock J. Physique {\bf 46}, 1843 (1985).

\bibitem{Nattermann+88a}
T.~Nattermann and J.~Villain,
\newblock Phase Transit. {\bf 11}, 5 (1988).

\bibitem{Patashinsky+79}
A.~Patashinsky and V.~Pokrovsky,
\newblock {\em Fluctuation Theory of Phase Transitions},
\newblock Pergamon Press, 1979.

\bibitem{Pokrovsky+79}
V.~L. Pokrovsky and A.~L. Talapov,
\newblock Phys. Rev. Lett. {\bf 42}, 65 (1979).

\bibitem{Horovitz+83}
B.~Horovitz, T.~Bohr, J.~M. Kosterlitz, and H.~J. Schulz,
\newblock Phys. Rev. B {\bf 28}, 6596 (1983).

\bibitem{Nozieres92}
P.~Nozi\`eres,
\newblock Shape and growth of crystals,
\newblock in {\em Solids far from Equilibrium}, edited by C.~Godr\`eche,
  Cambridge Univ. Press, Cambridge, 1992.

\bibitem{Huse+84}
D.~A. Huse and M.~E. Fisher,
\newblock Phys. Rev. B {\bf 29}, 239 (1984).

\bibitem{Blatter+94}
G.~Blatter, M.~V. Feigel'man, V.~B. Geshkenbein, A.~I. Larkin, and V.~M.
  Vinokur,
\newblock Rev. Mod. Phys. {\bf 66}, 1125 (1994).

\bibitem{Lemerle+98}
{S. Lemerle et al.},
\newblock Phys. Rev. Lett. {\bf 80}, 849 (1998).

\bibitem{Zaitsev+97}
S.~V. Zaitsev-Zotov, G.~Remenyi, and P.~Monceau,
\newblock Phys. Rev. Lett. {\bf 78}, 1098 (1997).

\bibitem{Alava+96}
M.~J. Alava and P.~M. Duxbury,
\newblock Phys. Rev. B {\bf 54}, 14990 (1996).

\bibitem{Sasada96}
T.~Sasada,
\newblock J. Phys. Soc. Japan {\bf 65}, 2889 (1996).

\bibitem{Emig+98b}
T.~Emig and T.~Nattermann,
\newblock unpublished .

\bibitem{Gingras+96}
M.~Gingras and D.~A. Huse,
\newblock Phys. Rev. B {\bf 53}, 15193 (1996).

\bibitem{Kierfeld+97}
J.~Kierfeld, T.~Nattermann, and T.~Hwa,
\newblock Phys. Rev. B {\bf 55}, 626 (1997).

\bibitem{FisherDS97}
D.~S. Fisher,
\newblock Phys. Rev. Lett. {\bf 78}, 1964 (1997).

\bibitem{Nattermann+89}
T.~Nattermann and P.~Rujan,
\newblock Int. J. Mod. Phys. B {\bf 3}, 1597 (1989).

\end{thebibliography}
\end{document}